\documentclass[12pt,a4paper]{article}

\usepackage{amsmath,amsfonts,latexsym,amssymb}
\usepackage{verbatim,amsthm,curves,graphics}
\usepackage{mathrsfs}

\theoremstyle{definition}
\newtheorem{thm}{Theorem}[section]

\newtheorem{lemma}{Lemma}[section]
\newtheorem{prop}{Proposition}[section]

\newtheorem{defn}{Definition}[section]

\begin{document}

\newcommand{\mye}{\stackrel{\circ}{e}}
\newcommand{\myn}{\stackrel{\circ}{\partial}}
\newcommand{\myI}{\stackrel{\circ}{I}}
\newcommand{\mr}{\mathbb{R}} 
\newcommand{\mz}{\mathbb{Z}} 
\newcommand{\mc}{\mathbb{C}} 
\newcommand{\mh}{\mathbb{H}} 
\newcommand{\ms}{\mathbb{S}} 
\newcommand{\mn}{\mathbb{N}} 
\newcommand{\F}{{\bf F}} 
\newcommand{\G}{{\bf G}} 
\newcommand{\pgator}{/\!\!\! S}
\newcommand{\rP}{\operatorname{P}}
\renewcommand{\i}{{\rm i}}
\newcommand{\car}{\operatorname{CAR}}
\newcommand{\rSpin}{\operatorname{Spin}}  
\newcommand{\rSO}{\operatorname{SO}}      
\newcommand{\rO}{\operatorname{O}}        
\newcommand{\rCliff}{\operatorname{Cliff}}
\newcommand{\WF}{\operatorname{WF}}       
\newcommand{\Pol}{{\rm WF}_{pol}}         
\newcommand{\cO}{\mathcal{O}}             
\newcommand{\cA}{\mathcal{A}}                
\newcommand{\cW}{\mathcal{W}}                
\newcommand{\fF}{\frak{A}}
\newcommand{\cV}{\mathscr{V}}             
\newcommand{\cP}{\mathcal{P}}             
\newcommand{\cC}{\mathcal{C}}             
\newcommand{\cF}{\mathscr{F}}             
\newcommand{\cE}{\mathcal{E}}             
\newcommand{\cH}{\mathscr{H}}
\newcommand{\cK}{\mathcal{K}}
\newcommand{\cM}{\mathscr{M}}
\newcommand{\cT}{\mathscr{T}}
\newcommand{\cN}{\mathscr{N}}
\newcommand{\cD}{\mathcal{D}}             
\newcommand{\cL}{\mathcal{L}}             
\newcommand{\cQ}{\mathscr{Q}}             
\newcommand{\cI}{\mathcal{I}}             
\newcommand{\bS}{\operatorname{S}}
\newcommand{\bU}{\mathcal{U}}
\newcommand{\bL}{\operatorname{OP}}
\newcommand{\glob}{{\textrm{\small global}}}
\newcommand{\loca}{{\textrm{\small local}}}
\newcommand{\singsupp}{\operatorname{singsupp}}
\newcommand{\dom}{\operatorname{dom}}
\newcommand{\clo}{\operatorname{clo}}
\newcommand{\supp}{\operatorname{supp}}
\newcommand{\rd}{{\rm d}}                 
\newcommand{\mslash}{/\!\!\!}             
\newcommand{\slom}{/\!\!\!\omega}         
\newcommand{\dirac}{/\!\!\!\nabla}        
\newcommand{\myid}{\leavevmode\hbox{\rm\small1\kern-3.8pt\normalsize1}}
\newcommand{\esssup}{\operatorname*{ess.sup}}
\newcommand{\ran}{\operatorname{ran}}
\newcommand{\sd}{{\rm sd}}
\newcommand{\reg}{\,{\rm l.n.o.}}
\newcommand{\wx}{{\bf x}}
\newcommand{\wk}{{\bf k}}
\newcommand{\ws}{{\bf s}}
\newcommand{\lno}{:\!}
\newcommand{\rno}{\!:}
\newcommand{\clim}{\operatorname*{coin.lim}}
\newcommand{\mydef}{\stackrel{\textrm{def}}{=}}
\renewcommand{\min}{{\textrm{\small int}\,\mathscr{L}}}
\renewcommand{\max}{\operatorname{max}}
\renewcommand{\Im}{\operatorname{Im}}
\renewcommand{\Re}{\operatorname{Re}}
\newcommand{\Exp}{\operatorname{Exp}}
\newcommand{\Ad}{\operatorname{Ad}}
\newcommand{\sa}{\mathfrak{sa} } 
\newcommand{\so}{\mathfrak{so} } 
\renewcommand{\o}{\mathfrak{o} } 
\newcommand{\su}{\mathfrak{su} } 
\newcommand{\sq}{\mathfrak{sq} } 
\renewcommand{\sp}{\mathfrak{sp} } 
\renewcommand{\sl}{\mathfrak{sl}} 
\newcommand{\gl}{\mathfrak{gl}} 
\newcommand{\g}{{\bf g} }	
\newcommand{\h}{\mathfrak{h} } 
\newcommand{\f}{\mathfrak{f} } 
\newcommand{\p}{\mathfrak{p} } 
\newcommand{\U}{\operatorname{U}} 
\newcommand{\Z}{\operatorname{Z}} 
\newcommand{\SO}{\operatorname{SO} } 
\newcommand{\SU}{\operatorname{SU} } 
\renewcommand{\O}{\operatorname{O} }
\newcommand{\SP}{\operatorname{Sp} } 
\newcommand{\SL}{\operatorname{SL}} 
\newcommand{\GL}{\operatorname{GL}} 
\newcommand{\Der}{\operatorname{Der}} 
\newcommand{\Str}{\operatorname{Str}} 
\newcommand{\End}{\operatorname{End}} 
\newcommand{\Cl}{\operatorname{Cl}} 
\newcommand{\ds}{\dot{+}}
\renewcommand{\aa}{\alpha}
\renewcommand{\gg}{\gamma}
\renewcommand{\d}{\delta }
\newcommand{\x}{\bar{x} } 
\newcommand{\y}{\bar{y} }
\newcommand{\s}{\bar{s} }
\newcommand{\diag}{\operatorname{diag}}
\newcommand{\ad}{\operatorname{ad}}
\newcommand{\sgn}{\operatorname{sgn}}
\newcommand{\tr}{\operatorname{tr}} 
\newcommand{\per}{\operatorname{per}} 
\newcommand{\Tri}{\operatorname{Tri}}

\title{Local Wick Polynomials and Time Ordered Products of Quantum Fields in Curved
       Spacetime}
\author{Stefan Hollands\thanks{Electronic mail: \tt stefan@bert.uchicago.edu}
        \,\, and
        Robert M. Wald\thanks{Electronic mail: \tt rmwa@midway.uchicago.edu}\\
       \it{Enrico Fermi Institute, Department of Physics,}\\
       \it{University of Chicago, 5640 Ellis Ave.,}\\
       \it{Chicago IL 60637, USA}
        }
\date{\today}

\maketitle

\begin{abstract}
In order to have well defined rules for the perturbative calculation
of quantities of interest in an interacting quantum field theory in
curved spacetime, it is necessary to construct Wick polynomials and
their time ordered products for the noninteracting theory. A
construction of these quantities has recently been given by Brunetti,
Fredenhagen, and K\"ohler, and by Brunetti and Fredenhagen, but they did
not impose any ``locality'' or ``covariance'' condition in their
constructions. As a consequence, their construction of time ordered
products contained ambiguities involving arbitrary functions of
spacetime point rather than arbitrary parameters. In this paper, we
construct an ``extended Wick polynomial algebra''---large enough to
contain the Wick polynomials and their time ordered products---by generalizing a construction of D\"utsch and Fredenhagen to curved spacetime. We then
define the notion of a {\it local, covariant quantum field}, and seek
a definition of {\it local} Wick polynomials and their time ordered
products as local, covariant quantum fields. We introduce a new notion
of the scaling behavior of a local, covariant quantum field, and
impose scaling requirements on our local Wick polynomials and their
time ordered products as well as certain additional
requirements---such as commutation relations with the free field and
appropriate continuity properties under variations of the spacetime
metric. For a given polynomial order in powers of the field, we prove
that these conditions uniquely determine the local Wick polynomials
and their time ordered products up to a finite number of
parameters. (These parameters correspond to the usual renormalization
ambiguities occurring in Minkowski spacetime together with additional
parameters corresponding to the coupling of the field to curvature.)
We also prove existence of local Wick polynomials. However, the issue
of existence of local time ordered products is deferred to a future
investigation. 
\end{abstract}

\pagebreak

\section{Introduction}

Despite some important differences from quantum field theory in
Minkowski spacetime caused by the lack of a ``preferred vacuum
state'', the theory of a linear quantum field in a globally
hyperbolic, curved spacetime is entirely well formulated (see, e.g.,
\cite{kw,w1} for a review). However, even in Minkowski spacetime, 
the theory of a nonlinear (i.e., self-interacting) quantum field is
not, in general, well formulated. Nevertheless, in Minkowski spacetime
there are well defined rules for obtaining perturbation series
expressions for all quantities of interest for a nonlinear field (and
in particular the interacting field itself). These perturbation
expressions are defined up to certain, well specified
``renormalization ambiguities''. It is of interest to know if a
similar perturbative definition of nonlinear quantum fields can be
given in curved spacetime and, if so, whether the renormalization
ambiguities in curved spacetime are of the same nature as those in
Minkowski spacetime.

This issue was analyzed by Bunch 
and collaborators \cite{bu,p} , but the key steps in this analysis were 
done in the
context of Riemannian spaces rather than Lorentzian spacetimes. Now,
Minkowski spacetime can be viewed as a real section of a complex
4-dimensional space that also contains a 4-dimensional, real Euclidean
section. It is well known that a suitable definition of a field theory
on this Euclidean section gives rise (via analytic continuation) to
the definition of a field theory in Minkowski spacetime. However, no
such connection between Riemannian and Lorentzian field theory holds
for curved spacetimes, since (apart from a few special classes of
spacetimes, such as static spacetimes) a general Lorentzian spacetime
cannot be expressed as a section of a complex spacetime that also
contains a real Riemannian section. Furthermore, the techniques used
by Bunch cannot readily be generalized to the Lorentzian case because
of the very significant mathematical differences in the nature of the
divergences occurring in the Riemannian and Lorentzian cases. For
example, in the Riemannian case, it follows from elliptic regularity
that Green's functions for the free theory are unique up to addition
of smooth functions. However, no such result holds in the Lorentzian case,
as exemplified by the very different properties of the advanced,
retarded, and Feynman propagators. Furthermore, singularities in the
Green's function occur only in the coincidence limit in the Riemannian
case, but they occur also for non-coincident, lightlike related events in
the Lorentzian case. As a result, formulas like (2.14) of \cite{bu}, which
play a crucial role in the Riemannian analysis, cannot be readily taken over
to the Lorentzian case. In addition, dimensional regularization and
other renormalization techniques used in Riemannian spaces are not well
defined in Lorentzian spacetimes.

Recently, significant progress in the definition of perturbative
quantum field theory in Lorentzian spacetimes was made by Brunetti and
Fredenhagen \cite{bfk,bf}, who used the methods of ``microlocal
analysis'' \cite{h,h1} 
to analyze the nature of the divergences occurring in the
Lorentzian theory. In \cite{bfk}, these authors considered the Fock space
arising (via the GNS construction) from a choice of quasi-free
Hadamard state $\omega$. They showed that on this Hilbert space, the
Wick polynomials---generated by the (formally infinite) products of
field operators and their derivatives evaluated at the same spacetime
point---can be given a well defined meaning as
operator-valued-distributions via a normal ordering prescription with
respect to $\omega$. In \cite{bf}, they then used an adaptation of the
Epstein-Glaser method \cite{eg} of renormalization in Minkowski spacetime to
analyze time ordered products of Wick polynomials, which are the
quantities needed for a perturbative construction of the interacting
field theory. They thereby showed that quantum field theories in
curved spacetime could be given the same ``perturbative
classification'' as in Minkowski spacetime, i.e., that all of the
``ultraviolet divergences'' of the theory in curved spacetime are of
the same nature as in Minkowski spacetime. Nevertheless, their
analysis in curved spacetime left open a much greater renormalization
ambiguity than in Minkowski spacetime: In essence, quantities that
appear at each perturbation order in Minkowski spacetime as
renormalized coupling {\it constants} now appear in curved spacetime
as renormalized coupling {\it functions}, whose dependence upon
the spacetime point can be arbitrary.

It seems clear that the missing ingredient in the analysis of \cite{bf}
is the imposition of a suitable requirement of covariance/locality on
the renormalization prescription, as was previously given for the
definition of the stress-energy tensor of a free quantum field (see
pp. 89--91 of \cite{w1}). The imposition of such a condition should provide an
appropriate replacement for the imposition of Poincare invariance in
Minkowski spacetime. When such a condition is imposed, one would
expect that the renormalized coupling functions would no longer be
arbitrary functions of the spacetime point but would be locally
constructed out of the metric in a covariant manner. Furthermore, one
might expect that when suitable continuity and scaling requirements
are also imposed, the ambiguities should be reduced to finitely many
free parameters at each order rather than free functions. The
renormalization ambiguities would then correspond to the
renormalization ambiguities in Minkowski spacetime together with the
renormalization of some additional parameters associated with
couplings of the quantum field to curvature.

The main purpose of this paper is to show that these expectations are
correct with regard to the uniqueness (though not necessarily 
the existence) of the perturbatively defined theory. 
A key step in our analysis is to define the notion of a {\it
local\footnote{In quantum field theory, the terminology 
``local field'' is commonly used to mean a field that commutes with 
itself at spacelike separated events. Our use of the terminology 
``local, covariant field'' here is not related to this notion. Rather, we
use this terminology to express the idea that the field is constructed
in a local and covariant way from the spacetime metric, as precisely
defined in Section \ref{sec2} below.}, 
covariant quantum field}. The basic idea behind this notion is
to consider a situation wherein one changes the metric outside of some
region $O$ and, in essence, demands that the local, covariant
quantum field not change within $O$.  A precise definition of this
notion will be given in Section
\ref{sec2} below (see Def. \ref{def2.1}). In Section \ref{sec2}, 
we will also explicitly see 
that the Wick polynomials as defined in \cite{bfk} fail to be local,
covariant quantum fields (no matter how $\omega$ is chosen);
consequently, neither are the time ordered products of these fields
constructed in
\cite{bf}. These quantities must therefore not be used for the
definition of the local observables in the interacting theory; their
definition depends on a choice of a reference state $\omega$, which is
itself a highly nonlocal quantity.

Our analysis will proceed as follows: First, we will obtain, for any
given globally hyperbolic spacetime $(M, \g)$, an abstract ``extended
Wick polynomial algebra'', $\cW(M, \g)$, via a normal ordering
prescription with respect to a quasi-free Hadamard state, $\omega$.
(We refer to our algebra $\cW(M, \g)$ as ``extended'', because it is
actually enlarged beyond the usual Wick polynomial algebra so as to
already include elements corresponding to the time ordered products of
Wick polynomials.)  Our construction of this algebra is essentially a
straightforward generalization to curved spacetime (using the methods
of \cite{bfk}) of a construction previously given by \cite{df} in the
context of Minkowski space.  We then note that the resulting operator
algebra---viewed as an abstract algebra---is independent of the choice
of $\omega$.  

Next, we will seek to identify the elements of this abstract algebra
that merit the interpretation of representing the various Wick
polynomials and time ordered products. As indicated above, the crucial  requirement that we shall
place on these elements is that they be local, covariant quantum
fields. We shall refer to
these elements as ``{\it local} Wick polynomials'' and ``{\it local}
time ordered products''. Some other ``specific properties''---such as
commutation relations with the free field---will also be imposed as
requirements on the definitions of these quantities. It is worth
emphasizing that, unlike in Minkowski space, we will find that some
ambiguities necessarily arise in defining local Wick
polynomials. Consequently, renormalization ambiguities in defining
perturbative quantum field theory in curved spacetime arise not only
from the definition of time ordered products of Wick polynomials but
also from the definition of the local Wick polynomials themselves.

As indicated above, after our locality/covariance requirement and our
other specific properties have been imposed on the definition of Wick
polynomials and their time ordered products, we will find that the ambiguities in the
definitions of these quantities will be reduced from arbitrary
functions of the spacetime point to functions that are locally constructed
from the metric (as well as parameters that appear in the classical
theory) in a covariant manner. However, in order to further reduce the
ambiguities to the renormalization of finitely many parameters at each
order, there are two other conditions we must impose: (i) a suitable
continuous/analytic dependence of the local Wick polynomials and their
time ordered products on the metric, $\g$, and coupling constants,
$p$, and (ii) a suitable scaling behavior of these
quantities. However, neither of these notions are straightforward to
define.

The difficulty with defining a suitable notion of the continuous
dependence of an element in $\cW(M, \g)$ 
on the metric and parameters occurring in the classical
theory arises from the fact that the Wick polynomial algebra $\cW(M, \g)$ 
for a spacetime $(M, \g)$ is not naturally isomorphic to the
Wick polynomial algebra $\cW(M, \g')$ for a different spacetime $(M,
\g')$, so it is far from clear what it means for an element of the
Wick polynomial algebra to vary continuously as $\g$ is
continuously varied to $\g'$.
Fortunately, the task of defining
this notion is made much easier by the fact that we are concerned only
with local, covariant quantum fields, so we may
restrict attention to metric variations that occur in some
spacetime region $O$ with compact closure. In order to make use of a similar
simplification with regard to variations of the parameters, $p$,
appearing in the classical theory, it is convenient to allow these
parameters to become functions of spacetime point and to then also
restrict attention to variations that occur only within $O$. If
$\g$ agrees with $\g'$ and $p$ agrees with $p'$ outside of
$O$, we can identify an element of $\cW_p(M, \g)$ with the element
of $\cW_{p'}(M, \g')$ which, say, agrees with it outside of
future\footnote{We would obtain a different identification of the
algebras by demanding agreement outside the past of $O$, but
this would give rise to an equivalent notion of continuous dependence.}
of $O$ (where we have a put a subscript $p$ 
on the algebras to indicate their dependence on the coupling parameters). 
With this identification of elements of the different
algebras, we require that if $(\g^{(s)}, p^{(s)})$
vary smoothly with $s$ in a suitable sense, then
within $O$ each local Wick polynomial and time ordered product
of local Wick polynomials must vary continuously with $s$. A precise
formulation of this requirement will be given in Section \ref{sec3.1} below.

The above requirement that the local Wick polynomials and their
time ordered products depend continuously on the metric would not
suffice to eliminate non-analytic local curvature ambiguities of the
sort considered in \cite{fla}. We therefore shall
impose an additional analyticity requirement that states that if
$\g^{(s)}$ is a one-parameter analytic family of analytic metrics,
then each local Wick polynomial and time ordered product of local Wick
polynomials must vary analytically with $s$; we similarly require
analytic variation of local Wick polynomials and their time ordered
products under analytic variation of the parameters $p$. However, for
analytic spacetimes, we cannot use the above method to identify 
algebras of different spacetimes, since one can no longer make local
variations of the metric. Instead, we proceed by introducing a notion
of an analytic family, $\omega^{(s)}$, of quasi-free Hadamard states on
$(M, \g^{(s)})$, and we require that the distributions obtained by acting
with $\omega^{(s)}$ on the local Wick polynomials and their
time ordered products vary analytically with $s$ in a suitable sense.
A precise formulation of these requirements will be given in Section \ref{sec3.1}.

In Minkowski spacetime, scaling behavior is usually formulated in
terms of how fields behave under the transformation $x \rightarrow
\lambda x$. Such a formulation would be highly coordinate
dependent in curved spacetime and thus would be very awkward to
implement. Our notion of local, covariant quantum fields allows us to
formulate a notion of scaling in terms of the behavior of these
fields under the scaling of the spacetime metric, $\g \rightarrow
\lambda^2\g$ (where $\lambda$ is a constant) together with associated 
scalings of the parameters, $p$, occurring in the theory.  Note that
in Minkowski spacetime, consideration of the behavior of a local,
covariant quantum field under scaling of the spacetime metric, $\g
\rightarrow
\lambda^2\g$, is equivalent to considering the behavior of these fields
under $x \rightarrow \lambda x$, since this diffeomorphism is a
conformal isometry with constant conformal factor $\lambda^{2}$, so $x
\rightarrow {\lambda} x$ with fixed metric is equivalent via a
diffeomorphism to $\g \rightarrow \lambda^{2} \g$ at each fixed
$x$. If we consider a classical field theory that is invariant under
$\g \rightarrow
\lambda^{2} \g$ together with corresponding scaling transformations
on the field and on the parameters, $p \rightarrow p(\lambda)$, 
appearing in the theory, then
the corresponding field algebras, $\cW_{p(\lambda)}(M, \lambda^{2}\g)$, will be
naturally isomorphic to each other. 
It might appear natural to require
that our definition of local Wick polynomials and their
time ordered products be such that they are preserved under this
isomorphism of the algebras. However, even in quantum field theory in
Minkowski spacetime, it is well known that such a requirement cannot
be imposed on time ordered products. In curved spacetime, we shall show that such a
scaling requirement cannot be imposed upon the local Wick polynomials
either. However, it is possible to require that the failure of the local
Wick polynomials and their time ordered products to scale like their
classical counterparts is given by terms with only logarithmic
dependence upon $\lambda$. This notion is made precise in Section
\ref{sec3.2}.

The main results of this paper may now be summarized. First, we shall
construct the algebra $\cW(M, \g)$ for an arbitrary globally
hyperbolic spacetime. We then define the notion of a ``local,
covariant quantum field'' and provide an axiomatic characterization of
``local Wick polynomials'' and their time ordered products. We shall then prove the existence of local
Wick polynomials via an explicit construction, and we shall give a
precise characterization of their non-uniqueness. Next, we consider
the time ordered products of local Wick polynomials. We shall obtain a
precise characterization of the non-uniqueness of these
time ordered products in a manner similar to our analysis of the
non-uniqueness of the local Wick polynomials. However, the existence
of time ordered products that satisfy our covariance/locality
requirement cannot be readily proven because the Epstein-Glaser
prescription does not manifestly preserve
covariance/locality. Consequently, we shall defer the investigation of
existence of time ordered products to a future investigation.

For simplicity and definiteness, we shall restrict consideration in this paper
to the theory of a real scalar field. However, the generalization of 
our definitions and conclusions to other fields should be 
straightforward. 

Notations and conventions: Throughout, $(M, \g)$ denotes a 
globally hyperbolic, time-oriented
spacetime. The manifold structure of $M$ is assumed to be real 
analytic, and the metric tensor $\g 
\equiv g_{ab}$ is assumed to be smooth (but not necessarily analytic). 
Our conventions regarding the spacetime geometry are those of \cite{w}. 
$V_x^\pm$ denote the closed future resp. 
past lightcone at a point $x$. 
$\square_\g = g^{ab}\nabla_a \nabla_b$ is the wave 
operator in curved space and $\mu_\g = |{\rm det} \,\g|^{1/2} d^4 x$. $\cD(M)$ is 
the space of (complex-valued) test functions with compact support on $M$ and 
$\cD'(M)$ is the corresponding dual space of distributions. Our 
convention for the Fourier transform in $\mr^n$ is 
$\hat u(k) = (2\pi)^{-n/2} \int e^{+\i kx} u(x) d^n x$. 

\pagebreak
\section{Definition of the extended Wick-polynomial algebra}

\subsection{Definition of the fundamental 
algebra of observables associated with a quantized 
Klein-Gordon field}

The theory of a free classical Klein-Gordon field on a spacetime $(M, \g)$ 
with mass $m$ and curvature coupling $\xi$ is described by the action 
\begin{equation}
\label{action}
{\cal S} =  \int_M \cL_0 \,\mu_\g = \int_M (g^{ab} \nabla_a \varphi \nabla_b \varphi +  
\xi R \varphi^2 + m^2 \varphi^2) \, \mu_\g. 
\end{equation}
The theory of a free quantized Klein-Gordon field in curved spacetime
can be formulated in various ways. For our purposes, it is 
essential to formulate the theory within the 
so-called ``algebraic approach'' (see, for example \cite{kw,w1}). 
In this approach, one starts from an abstract *-algebra 
$\cA(M, \g)$ (with unit), which is generated by certain 
expressions in the smeared quantum field, $\varphi(f)$, where $f$ is
a test function. In \cite{kw,w1}, expressions of the form $e^{\i\varphi(f)}$
were considered. The main advantage of working with such expressions is 
that the so-obtained algebra then has a norm (in technical terms, it is 
a $C^*$-algebra). Defining the algebra $\cA(M, \g)$ in that way would however be 
inconvenient for our purposes. 
Instead, we shall take $\cA(M, \g)$ to be the *-algebra generated by 
the identity and the smeared field operators 
$\varphi(f)$ themselves, subject to the following relations:

\medskip
\noindent
{\bf Linearity:} $\cD(M) \owns f \to \varphi(f) \in \cA(M, \g)$ is complex linear. \\
{\bf Klein-Gordon:} $\varphi((\square_\g - \xi R_\g - m^2)f)=0$ for all $f \in \cD(M)$.\\
{\bf Hermiticity:} $\varphi(f)^* = \varphi(\bar f)$. \\
{\bf Commutation Relations:}  $[\varphi(f_1), \varphi(f_2)] = \i\Delta_\g(f_1 \otimes f_2)1$, where
$\Delta_\g = \Delta^{\rm adv}_\g - \Delta^{\rm ret}_\g$ is the causal propagator 
for the Klein-Gordon operator. 

\medskip
\noindent
The so-obtained algebra 
$\cA(M, \g)$ is now no 
longer a $C^*$-algebra, because of the unbounded 
nature of the smeared quantum fields $\varphi(f)$. This will however not 
be relevant in the following.

A state in the algebraic framework is a linear functional 
$\omega: \cA(M, \g) \to \mc$ which is normalized so 
that $\omega(1) = 1$ and 
positive in the sense that $\omega(a^* a) \ge 0$ for all $a \in \cA(M, \g)$. 
The algebraic notion of a state is related to the usual 
Hilbert-space notion of a state by the GNS theorem. 
This says that for any algebraic state $\omega$, one can 
can construct a Hilbert space $\cH_\omega$
containing a distinguished ``vacuum''
vector $|\Omega_\omega \rangle$, and a representation 
$\pi_\omega$ of the algebraic elements $a \in \cA(M, \g)$ as 
linear operators on a dense invariant subspace $D_\omega \subset \cH_\omega$, 
such that $\omega(a) = 
\langle \Omega_\omega | \pi_\omega(a) | \Omega_\omega \rangle$ 
for all $a \in \cA(M, \g)$. The multilinear functionals on $\cD(M)$ defined
by 
\begin{equation}
\omega(f_1 \otimes \dots \otimes f_n) \mydef \omega( 
\varphi(f_1) \dots \varphi(f_n))
\end{equation}
are called $n$-point functions. 
Every state on $\cA(M, \g)$ is uniquely determined by the collection of its
$n$-point functions. A quasi-free state is by definition 
one which satisfies
\begin{equation}
\label{npoint}
\omega(e^{\i\varphi(f)}) = e^{-\frac{1}{2} \omega(f \otimes f)}. 
\end{equation}
Note that the elements $e^{\i\varphi(f)}$ do not actually belong to 
the algebra $\cA(M, \g)$. What is meant by Eq. \eqref{npoint} is 
the set of identities obtained by functionally differentiating 
this equation with respect to $f$. The so obtained identities then 
express the $n$-point functions of the state $\omega$ in 
terms of its two-point function.
For quasi-free states, the GNS construction gives the usually considered
representation of the fields on Fock-space, with $|\Omega_\omega \rangle$
the Fock-vacuum and with the field given in terms of creation and 
annihilation operators \cite{kw}. 

In our subsequent constructions, we will consider quasi-free states which 
are in addition of ``global Hadamard type''. These are states 
whose two-point function has no spacelike singularities, and whose 
symmetrized two-point functions is given locally, modulo a smooth 
function, by a Hadamard fundamental solution \cite{dw}, $H$, defined as 
\begin{equation}\label{Hdef}
H(x, y) = u(x, y)\,{\rm P} (\sigma^{-1}) + v(x, y) \ln |\sigma|. 
\end{equation} 
Here, $\sigma$ is the squared geodesic distance between the points 
$x$ and $y$ in the spacetime $(M, \g)$, $u$ and $v$ are certain
real and symmetric smooth functions constructed from the metric and 
the couplings and ``P'' denotes the principal value. Strictly speaking, $H$ is 
well defined only in analytic spacetimes (we will come back to this issue in 
Sec. \ref{sec5.2}), so the above definition needs to be 
modified in spacetimes that are only smooth. 
For a detailed discussion of this and of the statement that ``there are no spacelike
singularities'', see \cite{kw}. 
An immediate consequence the definition of Hadamard 
states is that if $\omega$ and $\omega'$ are 
Hadamard states, then $\omega(x, y) - \omega'(x, y)$ is a smooth function
on $M \times M$. 

There exists an alternative, equivalent characterization of globally Hadamard 
due to Radzikowski \cite[Thm. 5.1]{r}, involving the notion of the ``wave front set''\cite{h, h1} 
of a distribution, which will play a crucial role in our subsequent 
constructions. (A definition of the wave front set and some of 
its elementary properties is given in the Appendix.) 
Namely, the globally Hadamard states in the sense of \cite{kw} are precisely 
those states whose two-point function is a bidistribution with wave front set 
\begin{eqnarray}
\label{wfs}
\WF(\omega) = \{ (x_1, k_1, x_2, -k_2) \in (T^*M)^2 \backslash \{0\}  
\mid (x_1, k_1) \sim (x_2, k_2), k_1 \in V_{x_1}^+ \}.
\end{eqnarray}
Here, the following notation has been used: We write $(x_1, k_1) \sim (x_2, k_2)$ if 
$x_1$ and $x_2$ can be joined by a null geodesic and if $k_{1}$ and $k_{2}$ are
cotangent and coparallel to that null geodesic. 

\subsection{Definition and properties of the algebra $\cW(M, \g)$}
\label{sec2.2}

In the previous subsection, we reviewed the algebraic construction of 
a free quantum field theory. However, the algebra $\cA(M, \g)$ used in 
that construction includes only observables corresponding to the
smeared $n$-point functions of the free field. If we wish
to define a nonlinear quantum field theory via a perturbative 
construction off the free field theory, we must consider 
additional observables, namely Wick polynomials and their
time ordered products. Our strategy for doing so is to 
define an enlarged algebra of observables, $\cW(M, \g)$, that
contains $\cA(M, \g)$ and also contains, among others, elements 
corresponding to 
(smeared) Wick polynomials of free-fields and (smeared) 
time ordered products of these fields. The
construction of $\cW(M, \g)$ is essentially a straightforward generalization of \cite{df}, 
using ideas of \cite{bfk, bf}. The construction initially depends on 
the choice of an arbitrary quasi-free Hadamard state $\omega$ on $\cA(M, \g)$. 
However, we will show below that different choices for 
$\omega$ give rise to isomorphic algebras. In that sense the algebras $\cW(M, \g)$ do not depend 
on the choice of a particular quasi-free Hadamard state. 
We note that, in particular, the construction of $\cW(M, \g)$ achieves the goal
stated on p.~86 of \cite{w1}, namely, to define an enlarged algebra of 
observables that includes the smeared stress-energy tensor.

Once we have properly identified the elements in $\cW(M, \g)$ 
corresponding to local Wick products and local time ordered
products, the standard rules of perturbative quantum field
theory will allow us to obtain perturbative expressions for 
the interacting field observables. These perturbative 
quantities---such as for example the interacting field itself---are 
given by formal power series in the coupling constants. 
The infinite sums occurring in these formal power series do not, of course, define
elements of our algebra $\cW(M, \g)$.
However, the expressions obtained by truncating these power series 
at some arbitrary order in perturbation theory
will be elements in $\cW(M, \g)$. In that sense $\cW(M, \g)$ 
contains the observables (to arbitrary high order in perturbation theory) 
of the interacting theory. The ``renormalization ambiguities'' occurring in these
perturbative expressions arise from the ambiguities in the definition of the local 
Wick products and local time ordered products. The main goal in 
this paper is to give a precise characterization of these ambiguities. 

It should be noted that 
since $\cA(M, \g) \subset \cW(M, \g)$, the notion of states for the 
nonlinear theory will be more restrictive than the notion of 
states for the free theory given in the previous section, but the 
states on $\cW(M, \g)$ will include a dense set of vectors in 
the GNS representation of any quasi-free Hadamard state. Indeed, it 
will follow from our results below that all Hadamard states on $\cA(M, \g)$
whose truncated $n$-point functions (other than the two-point 
function) are smooth can be extended to $\cW(M, \g)$. We 
conjecture that these are the only states on $\cA(M, \g)$ that 
can be extended to $\cW(M, \g)$, i.e., that the states on 
$\cW(M, \g)$ are in 1--1 correspondence with Hadamard states on 
$\cA(M, \g)$ with smooth truncated $n$-point functions.\footnote{
Kay (unpublished) has shown that in the vacuum representation 
of $\cA$ in Minkowski spacetime, these states include all
$n$-particle states with smooth mode functions. More generally, he
also showed that on a globally hyperbolic 
spacetime, these states include all 
$n$-particle states with smooth mode functions in the 
GNS representation of any quasi-free Hadamard state. 
}

\medskip

To begin our construction of $\cW(M, \g)$, choose a quasi-free 
Hadamard state $\omega$ on $\cA(M, \g)$. Via 
the GNS construction, one obtains from this a representation of 
the field operators $\varphi(f)$ as linear operators on a  
Hilbert space $\cH_\omega$ with dense, invariant domain $D_\omega$, where
we use the same symbol for the algebraic element 
$\varphi(f)$ and its representative on $\cH_\omega$. 
Next, define the symmetric operator-valued distributions
\begin{multline}
\label{wndef}
W_n(x_1, \dots, x_n) = \; \lno \varphi(x_1) \dots \varphi(x_n) \rno_\omega \; \mydef \\  
\frac{\delta^n}{\i^n \delta f(x_1) \dots \delta f(x_n)} {\rm exp}\left[
\frac{1}{2}\omega(f \otimes f) + \i\varphi(f)\right] \Bigg|_{f = 0}
\end{multline}
for $n \ge 1$ and $W_0 \equiv 1$. 
The operators $W_n(t)$ obtained by smearing with a test function $t = 
f_1 \otimes \dots \otimes f_n \in \cD(M^n)$ are elements the algebra $\cA(M, \g)$. 
The product of two operators $W_n(t)$ and $W_m(t')$ is given 
by the following formula (which is just a re-formulation of Wick's theorem), 
\begin{eqnarray}
\label{wprd}
W_n(t)W_m(t') = \sum_{k}W_{n+m-2k}(t \otimes_{k} t') \quad \forall t \in \cD(M^n), \; t' \in \cD(M^m). 
\end{eqnarray}
The expression $t \otimes_k t'$ is the symmetrized, $k$ times contracted tensor product, 
defined for $m, n \ge k$ by 
\begin{multline}
\label{conpr}
(t \otimes_{k} t')(x_1, \dots, x_{n+m-2k}) \mydef 
{\bf S} \frac{n!m!}{(n-k)!(m-k)!k!}
\int_{M^{2k}} 
t(y_{1}, \dots, y_{k}, x_1, \dots, x_{n-k})\times\\
t'(y_{k+1}, \dots, y_{k+i}, x_{n-k+1}, 
\dots, x_{n+m-2k})\prod_{i=1}^k \omega(y_{i}, y_{k+i}) \,
\mu_\g(y_{i}) \mu_\g(y_{k+i}) 
\end{multline} 
where $\bf S$ means symmetrization in $x_1, \dots, x_{n+m-2k}$. If either 
$m < k$ or $n < k$, then the contracted tensor product is defined to be 
zero. 

\medskip

In order to obtain more general operators such as normal ordered Wick powers, 
we would like to be able to smear the operator-valued distributions $W_n$ 
not only with smooth test {\it functions}, but in addition 
also with certain compactly supported test {\it distributions} $t$. That this 
is indeed possible can be seen by means of a microlocal argument, which is based 
on the following observation \cite{bf}: The domain $D_\omega$ 
contains a dense invariant subspace of vectors $|\psi\rangle$ 
(the so-called ``microlocal domain of smoothness'', see
\cite[Eq. (11)]{bf})
having the property that the wave front set 
of the vector-valued distributions $t \to W_n(t)|\psi\rangle$ 
is contained in the set $\F_n(M, \g)$, defined as 
\begin{eqnarray}
\F_n(M, \g) =  \{ (x_1, k_1, \dots, x_n, k_n) \in (T^*M)^n \backslash \{0\}
\mid k_i \in V^-_{x_i}, i = 1, \dots, n \}.
\end{eqnarray}
Now, smearing the above vector-valued distributions with a distributional test function
$t$ involves taking the pointwise product of two distributions. As it is well known,  
the pointwise product of two distributions is in general ill-defined. However, 
a theorem by H\"ormander \cite[Thm. 8.2.10]{h} states that if the wave front sets 
of two distributions $u$ and $v$ are such that $\{0\} \notin \WF(u) 
+ \WF(v)$, then the pointwise product between $u$ and $v$ {\it can} be unambiguously 
defined. In the case at hand, we are thus allowed to smear $W_n$ in with any 
compactly supported distribution $t$ such that 
$\{0\} \notin \WF(t) + \F_n(M, \g)$. We here shall consider a subclass 
of the set of all such $n$-point distributions $t$, namely the class
\begin{eqnarray}
\cE'_n(M, \g) \mydef \{ t \in \cD'(M^n) \mid \text{$t$ is symmetric, $\supp(t)$ is compact, 
$\WF(t) \subset \G_n(M, \g)$} 
\}, 
\end{eqnarray}
where 
\begin{equation}
\G_n(M, \g) \mydef (T^*M)^n \backslash \left(
\bigcup_{x \in M} (V^+_x)^n \cup  
\bigcup_{x \in M} (V^-_x)^n \right).
\end{equation}
Smearing $W_n$ with test distributions $t \in \cE'_n(M, \g)$ gives therefore 
well defined operators on the microlocal domain 
of smoothness. (For notational simplicity, 
we denote this domain again by $D_\omega$.)
\begin{defn}
$\cW(M, \g)$ is the *-algebra of operators on $\cH_\omega$ 
generated by $1$ and elements of the form 
$W_n(t)$, where $n \ge 1$ and where $t \in \cE'_n(M, \g)$. 
\end{defn}

\begin{thm}
\label{1.1}
The product in the algebra $\cW(M, \g)$ can be 
computed by Eq. \eqref{wprd}, and the *-operation is given 
by $W_n(t)^* = W_n(\bar t)$. Furthermore, $W_n(t)=0$ whenever
$t$ is of the form $t(x_1, \dots, x_n) = (\square_\g - \xi R_\g - m^2)_{x_i}
s(x_1, \dots, x_n)$ for some $s \in \cE'_n(M, \g)$.  
\end{thm} 
\begin{proof}
The statement concerning the *-operation is obvious.
In order to show that the algebra product can be 
calculated by Eq. \eqref{wprd}, we first show that if  
$t \in \cE'_n(M, \g)$ and $t' \in \cE_m'(M, \g)$, 
then $t \otimes_k t' \in 
\cE'_{n+m-2k}(M, \g)$. Clearly $t \otimes_k t'$ is compactly supported and 
symmetric. We must show
that in addition $\WF(t \otimes_k t') \subset {\bf G}_{n+m-2k}(M, \g)$. 
This can be seen by an application of \cite[Thm. 8.2.13]{h}, which yields, 
in combination with Eq. \eqref{wfs} for $\WF(\omega)$, 
\begin{eqnarray}
\label{ttprk}
\WF(t \otimes_k t') &\subset& 
\{ (x_1, k_1, \dots, x_{n+m-2k}, k_{n+m-2k}) \in 
(T^*M)^{n+m-2k} \mid \nonumber\\
&& \text{$\exists$ elements $(x_1, k_1, \dots, x_{n-k}, k_{n-k}, y_1, p_1, \dots, y_k, 
p_k)
\in \WF(t)$ and} \nonumber \\
&&\text{$(x_{n-k+1}, k_{n-k+1}, \dots, x_{n+m-2k}, k_{n+m-2k},
y_{k + 1}, p_{k + 1}, \dots, y_{2k}, p_{2k}) \in \WF(t')$} 
\nonumber\\
&&\text{such that either $(x_j, p_j) \sim  (x_{j+k}, -p_{j+k})$ and  
$p_j \in V_{x_j}^- \backslash \{0\}$}
\nonumber\\
&&\text{or $p_j = p_{j+k} = 0$ for all $j=1, \dots, k$}\}.    
\end{eqnarray}
It is not difficult to see that the set on the 
right side of the above inclusion is in fact contained in 
$\G_{n+m-2k}(M, \g)$, thereby showing that 
$t \otimes_k t'$ is in the class $\cE'_{n+m-2k}(M, \g)$, 
as we wanted to show. We finish the proof by showing that Eq. \eqref{wprd} holds not
only for smooth test functions, but also for our admissible test distributions
$t \in \cE'_n(M, \g)$ and $t' \in \cE'_m(M, \g)$. 
To see this, we consider sequences of test functions
$\{t_\alpha\}$ and $\{t'_\alpha\}$ converging to 
$t$ and $t'$ in the sense of $\cD'_{\Gamma_n}(M^n)$ resp. $\cD'_{\Gamma_m}
(M^m)$ (for a definition of these spaces and their pseudo topology, the so-called 
``H\"ormander pseudo topology'', see the Appendix), where $\Gamma_n$ and
$\Gamma_m$ are closed conic sets in $\G_n(M, \g)$ and
$\G_m(M, \g)$, respectively with the property that 
$\WF(t) \subset \Gamma_n$ and $\WF(t') \subset \Gamma_m$. 
Now the operation of composing distributions---which forms the basis of 
the definition of the contracted tensor product, Eq. \eqref{conpr}---is 
continuous in the H\"ormander pseudo topology. Therefore
$t_\alpha \otimes_k t'_\alpha \to 
t \otimes_k t'$ in the space $\cD'_{\Gamma_{m+n-2k}}(M^{n+m-2k})$, 
where  $\Gamma_{n+m-2k}$
is a certain closed conic set in $\G_{n+m-2k}(M, \g)$, which is 
calculable from $\Gamma_n$ and $\Gamma_m$ using formula 
Eq. \eqref{ttprk}.  

Now expressions of the sort $W_n(t)|\psi\rangle$ arise
from the pointwise product of distributions. This product is continuous 
in the H\"ormander pseudo topology. Therefore we conclude that 
$W_{n+m-2k}(t_\alpha \otimes_k t'_\alpha)
|\psi\rangle \to W_{n+m-2k}(t \otimes_k t')
|\psi\rangle$. By a similar argument, it also follows that 
$W_n(t_\alpha)W_m(t'_\alpha) |\psi\rangle \to W_n(t)W_m(t') |\psi\rangle$.    
Eq. \eqref{wprd}, applied to some vector $|\psi\rangle \in D_\omega$, 
is already known to hold for $t_\alpha$ and $t'_\alpha$, 
since these are smooth test functions. It follows that Eq. \eqref{wprd} must also hold 
for our admissible test distributions. 

The last statement of the 
theorem is obvious from the definition of $W_n$ when $t$ and 
$s$ are smooth functions. By a continuity argument similar to the one
above, it also holds for distributional $t$ and $s$. 
\end{proof}

Since $\cE'_n(M, \g)$ is a vector space and since Eq. \eqref{wprd} holds, it 
follows immediately that any $a \in \cW(M, \g)$ can be written in the 
form
\begin{eqnarray}
\label{aform}
a = t_0 1 + \sum_{n=1}^N W_n(t_n),
\end{eqnarray}
with $t_0 \in \mc$ and $t_n \in \cE_n'(M, \g)$. Furthermore, the following
proposition holds, which will be needed in Sec. \ref{sec5}: 
\begin{prop}
\label{pro1}
Let $k \ge 0$ and let
$a \in \cW(M, \g)$ be such that 
\begin{equation}
[ \dots [[a, \varphi(f_1)], \varphi(f_2)], \dots \varphi(f_{k+1})] = 0 
\quad \forall  f_1, \dots, f_{k+1} \in \cD(M).
\end{equation}
Then $a$ is of the form
$a = t_0 1 + \sum_{n=1}^k W_n(t_n)$, where $t_0 \in \mc$ and $t_n \in \cE'_n(M, \g)$.   
\end{prop} 
\begin{proof}
$a$ must be of the form \eqref{aform} where $N$ is some natural number. We must show
that $N \le k$. Let us assume that $N > k$ and that $W_N(t_N) \neq 0$. 
We show that this leads to a contradiction. By
assumption $[ \dots [a, \varphi(f_1)], \dots \varphi(f_{N+1})] = 0$ for all 
test functions. Using Eq. \eqref{wprd} (and recalling that $\varphi(f) = 
W_1(f)$), this gives us 
\begin{equation}
\label{dtg}
(\Delta_\g \otimes \dots \otimes \Delta_\g) t_N(x_1, \dots, x_N) \equiv 0. 
\end{equation}
Using the relation $\Delta_\g = \Delta^{\rm adv}_\g - \Delta_\g^{\rm ret}$, the support 
properties of the advanced and retarded fundamental solutions and the fact 
that $t_N$ is compactly supported,  
one finds from Eq. \eqref{dtg} that the distribution 
$s = (\Delta^{\rm ret}_\g \otimes \dots \otimes \Delta^{\rm ret}_\g) t_N$ 
must be of compact support. In combination with a microlocal argument similar to the
one given in the proof of Thm. \ref{1.1}, one finds moreover that $s \in \cE'_N(M, \g)$. Since  
$t_N(x_1, \dots, x_N) = \prod_{i=1}^N (\square_\g - \xi R_\g - m^2)_{x_i} s(x_1, \dots, x_N)$, 
it follows from Thm. \ref{1.1} that $W_N(t_N) = 0$, which contradicts our 
hypothesis.
\end{proof}

That the algebra $\cW(M, \g)$ contains normal ordered Wick products can be seen as follows. Let  
\begin{equation}
\label{tdef}
t(x_1, \dots, x_k) = f(x_1) \delta(x_1, \dots, x_k),
\quad f \in \cD(M).
\end{equation}
The distribution $t$ is in $\cE_k'(M, \g)$, because  
$$\WF(t) = \{ (x, k_1, \dots, x, k_k) \in (T^*M)^k \backslash \{0\}\mid
\sum_i k_i = 0 \} \subset \G_k(M, \g).$$
The algebraic element $W_k(t)$ with $t$ as in Eq. \eqref{tdef} is then 
just the $n$-th normal ordered Wick power of a free field operator, as 
previously defined in \cite{bfk},  
\begin{equation}
\label{nodef}
\lno \varphi^k (f)\rno_{\omega} \; = W_k(t).
\end{equation} 
More generally, we may take $t$ to be 
\begin{eqnarray}
t(x_1, \dots, x_r) = \delta(x_{i_1}, \dots) \dots \delta(x_{i_n}, \dots) f_1(x_{i_1}) \dots f_r(x_{i_n})
\end{eqnarray} 
where $I_1 = \{i_1, \dots \}, \dots, I_n = \{i_n, \dots \}$ is a partition of $\{1, \dots, r\}$
into $n$ pairwise disjoint subsets with $|I_j| = k_j$. This gives us the generalized Wick product
\begin{equation}
\label{nodef1}
\lno \varphi^{k_1} (f_1) \dots \varphi^{k_n}(f_n)\rno_{\omega} \; = W_r(t).
\end{equation} 
As was shown in \cite{bf},  
$\cW(M, \g)$ also contains time ordered products of Wick-powers 
of free fields. 

\medskip

We next discuss the dependence of the algebra $\cW(M, \g)$ on our choice of a 
reference state $\omega$. Let us suppose we had started with another 
quasi-free Hadamard state $\omega'$. We would then have obtained another algebra
$\cW'(M, \g)$ generated by corresponding operators
acting on the GNS Hilbert space constructed from $\omega'$.
If the GNS representations of $\omega$ and $\omega'$ were unitarily 
equivalent, then the Bogoliubov transformation implementing that unitary 
equivalence would induce a canonical isomorphism between $\cW(M, \g)$ and
$\cW'(M, \g)$. However, even if the GNS representations of $\omega$ and
$\omega'$ fail to be unitarily equivalent, at the algebraic level, 
there is nevertheless a canonical isomorphism: 
\begin{lemma}
\label{ilemma}
There is a canonical *-isomorphism 
$\alpha: \cW'(M, \g) \to \cW(M, \g)$, which acts on the generators 
$W_n'$ of $\cW'(M, \g)$ by  
\begin{eqnarray}
\label{idef}
\alpha(W_n'(t)) \mydef \sum_{k}
W_{n - 2k}(\langle d^{\otimes k}, t\rangle),   
\end{eqnarray}
where $W_n$ denote the generators in $\cW(M, \g)$, and we 
are using the following notation: $d(x_1, x_2) = \omega(x_1, x_2) - 
\omega'(x_1, x_2)$ and 
\begin{multline}
\label{intdef}
\langle d^{\otimes k}, t \rangle
(x_1, \dots, x_{n-2k}) \mydef \frac{n!}{(2k)!(n-2k)!}
\int_{M^{2k}} 
t(y_1, \dots, y_{2k}, x_1, \dots, x_{n-2k}) \times\\
\prod_{i = 1}^k d(y_{2i-1}, y_{2i})  \,\mu_\g(y_{2i-1})
\mu_\g(y_{2i}) 
\end{multline}
for $2k \le n$ and $\langle d^{\otimes k}, t \rangle = 0$ for $2k > n$.
\end{lemma}
\begin{proof}
In order to show that the right hand side of Eq. \eqref{idef}
represents an element in $\cW(M, \g)$,  
we must show that $\langle d^{\otimes k}, t \rangle \in \cE'_{n-2k}(M, \g)$. 
We first note that, since $\omega$ and $\omega'$ are Hadamard states, 
$d$ is smooth. By \cite[Thm. 8.2.13]{h} we therefore find  
\begin{multline}
\WF(\langle d^{\otimes k}, t \rangle) \subset
 \{ (x_1, k_1, \dots, x_{n-2k}, k_{n-2k}) \in 
(T^*M)^{n-2k} \backslash \{0\} \mid \\
\exists (x_1, k_1, \dots, x_{n-2k}, k_{n-2k}, y_1, 0, \dots, y_{2k}, 0) \in
\G_n(M, \g) \} \subset \G_{n-2k}(M, \g).
\end{multline} 
The distribution $\langle d^{\otimes k}, t \rangle$ is by 
definition symmetric and of compact support. 
Therefore $\langle d^{\otimes k}, t \rangle \in \cE'_n(M, \g)$, 
which gives us that $\alpha(W_n(t)) \in \cW(M, \g)$.  
Since every element in $\cW'(M, \g)$ can be written as a 
sum of elements of the form $W'_n(t)$, with $t \in \cE'_n(M, \g)$, 
we may therefore take Eq. \eqref{idef} as the definition of a linear 
map from $\cW'(M, \g)$ to $\cW(M, \g)$. 
That this map is a homomorphism is demonstrated by the 
following calculation.  
\begin{eqnarray}
\alpha(W'_n(t))\alpha(W'_m(t')) &=& \sum_{k,l}
W_{n-2k}(\langle d^{\otimes k}, t \rangle) 
W_{n-2l}(\langle d^{\otimes l}, t' \rangle) \nonumber\\
&=& \sum_i\sum_{k,l}
W_{n+m-2(k+l+i)}(\langle d^{\otimes k}, t \rangle
\otimes_i \langle d^{\otimes l}, t' \rangle) \nonumber\\
&=& \sum_i\sum_{r}\sum_{k = 0}^r
W_{n+m-2(r+i)}(\langle d^{\otimes k}, t \rangle
\otimes_i \langle d^{\otimes (r-k)}, t' \rangle) \nonumber\\
&=& \sum_i\sum_{r}
W_{n+m-2(r+i)}(\langle d^{\otimes r}, t \otimes_i t' \rangle) 
\nonumber\\
&=& \alpha(W'_n(t)W'_m(t')),  
\end{eqnarray}
where we have used the identity 
\begin{equation}
\sum_{k = 0}^r \langle d^{\otimes k}, t \rangle
\otimes_i \langle d^{\otimes (r-k)}, t' \rangle = 
\langle d^{\otimes r},
t \otimes_i t' \rangle.
\end{equation}
That $\alpha$ preserves the *-operation follows because $d$ is real, 
which is in turn a consequence of the fact that ${\rm Im\,} \omega = {\rm Im\,} \omega' = 
\frac{1}{2} \Delta_\g$. 
That $\alpha$ is one-to-one can be
seen from an explicit construction of its inverse, given by 
the same formula as \eqref{idef}, but with $d$ replaced by $-d$. 
\end{proof}

It should be noted here that the abstract algebra $\cW(M, \g)$ could be 
defined more simply and directly as the algebra of expressions of 
the form Eq. \eqref{aform}, with a product defined by Eq. \eqref{wprd}, 
a *-operation defined by $W_n(t)^* = W_n(\bar t)$ and which satisfy 
$W_n(t) = 0$ whenever $t$ is of the form $t(x_1, \dots, x_n) = 
(\square_\g - \xi R_\g - m^2)_{x_i} s(x_1, \dots, x_n)$. (Note, however, that the
definition of the product \eqref{wprd} requires a choice of Hadamard 
state $\omega$; see Eq. \eqref{conpr}.) However, our explicit 
construction of $\cW(M, \g)$ as an operator algebra on the GNS representation 
of a quasi-free state, $\omega$, on $\cA(M, \g)$, is useful for establishing
that a suitably wide class of states exists on $\cW(M, \g)$.
In addition, the concrete realization of $\cW(M, \g)$ will be 
useful in our explicit construction of local Wick products. 

\medskip

For later purposes, we also need to define a notion of convergence 
within the algebra $\cW(M, \g)$.
In particular, we would like to have a notion of convergence which is 
preserved under taking products in our algebra, and which is 
independent of the quasi-free
Hadamard state $\omega$ by which this algebra is defined.  
Such a notion can be defined as follows.

Let $\{t_\alpha\}$ be a sequence of distributions in $\cE_n'(M, \g)$
with $\WF(t_\alpha) \subset \Gamma_n \; \forall \alpha$, where
$\Gamma_n$ is some closed conic set contained in $\G_n(M, \g)$. 
Then we say that 
$$a_\alpha = W_n(t_\alpha) \to 
a = W_n({t})
\quad \text{in $\cW(M, \g)$} 
$$
if 
$$
\text{$t_\alpha \to t$ 
in $\cD'_{\Gamma_n}(M^n)$,}
$$
i.e., if $t_\alpha \to t$ in the sense of the 
H\"ormander pseudo topology associated with the cone $\Gamma_n$
(for the definition of this pseudo topology and the spaces 
$\cD'_{\Gamma_n}(M^n)$ we refer to the Appendix).
Convergence in the H\"ormander pseudo topology guarantees that 
$t \in \cE_n'(M, \g)$. Therefore our algebra is closed with respect to 
the above notion of convergence. Clearly, that notion is also independent of 
the particular quasi-free Hadamard state chosen to define $\cW(M, \g)$. 
Finally, let $a_\alpha \to a$ and $b_\alpha \to b$ be two convergent sequences in 
$\cW(M, \g)$ in the above sense.  Then, by an argument almost identical 
to the one given towards the end of the proof of Thm. \ref{1.1}, 
we also have $a_\alpha b_\alpha \to ab$. 
Hence, the element-wise product of two convergent sequences of algebraic elements 
gives again a convergent sequence. 

\section{Mathematical formulation of the notion of a local, covariant 
quantum field}
\label{sec2}
The field quantities of interest in quantum field theory 
in curved spacetime such as the stress energy tensor of 
free fields or the quantity ``$\lambda \varphi^4$'' should be
local and covariant, i.e., their definition should not depend
on structures that are only globally defined (such as a 
preferred vacuum state) nor should they depend on non-covariant
structures (such as a preferred coordinate system).
The aim of this section is to explain precisely what 
we mean by the statement that 
an element in $\cW(M, \g)$ is ``locally defined'' and 
``transforms covariantly under diffeomorphisms''. 
This notion requires the consideration
of a given operator on spacetimes $(M, \g)$ and $(M', \g')$ 
that have isometric regions, but that are not 
globally isometric. The basic problem is 
that operators living on $(M, \g)$ and $(M', \g')$ belong to different 
algebras, and therefore cannot be compared directly. Therefore, we must first 
provide a natural and consistent identification of the corresponding algebras
(see Lem. \ref{2.1}).
For this purpose, we consider ``causality preserving
isometric embeddings'', that is, isometric embeddings $\chi: 
N \to M$ from a spacetime $(N, \g')$ to another spacetime $(M, \g)$
so that the causal structure on $\chi(N)$ induced from $(N, \g')$ 
coincides with that induced from $(M, \g)$. (This is equivalent
to the condition that $\chi$ preserves the time-orientation and 
that $J^+(x) \cap J^-(y) \subset \chi(N) \,\, \forall x, y \in \chi(N)$.) 

\begin{lemma}
\label{2.1}
Let $\chi: N \to M$ be an isometric embedding of some globally
hyperbolic spacetime $(N, \g')$ into another globally hyperbolic
spacetime $(M, \g)$
(so that in fact $\g' = \chi^* \g$) which is causality 
preserving. Denote by $\cW(N, \g')$ and 
$\cW(M, \g)$ the corresponding extended Wick-polynomial algebras, 
viewed as abstract algebras. 
Then there is a natural injective *-homomorphism  
$\iota_\chi: \cW(N, \g') \to \cW(M, \g)$ such that if 
$\omega$ is a quasi-free Hadamard state on $(M, \g)$ and 
$\omega'(x, y) = \omega(\chi(x), \chi(y))$ we have 
\begin{eqnarray}
\iota_\chi(W_n'(t)) =
W_n(t \circ \chi^{-1}) \quad \forall t \in \cE_n'(N, \g'), 
\end{eqnarray}  
where $W'_n$ and $W_n$ are given by Eq. \eqref{wndef} in the GNS representations 
of $\omega'$ and $\omega$ respectively and 
$\chi^{-1}: \chi(N) \to N$ is the inverse of 
$\chi$ (defined on the image of $N$ under $\chi$).
\end{lemma}
\begin{proof}
Let $\omega$ be a quasi-free Hadamard state for the spacetime 
$(M, \g)$ and let $\omega'(x, y) = \omega(\chi(x), \chi(y))$. 
Then $\omega'(x, y)$ is the two-point function of a quasi-free
Hadamard state $\omega'$ on $(N, \g')$. (Here we are 
using the assumption that our isometry $\chi$ is causality preserving.)  
By Lem. \ref{ilemma}, 
we may assume that the abstract algebras $\cW(N, \g')$ 
and $\cW(M, \g)$ are concretely realized as linear operators on 
the GNS constructions of the quasi-free Hadamard states 
$\omega'$ and $\omega$. Since every element in 
$\cW(N, \g')$ can be written as a sum of elements of 
the form $W_n(t)$, the above formula gives, by linearity, 
a map from $\cW(N, \g')$ to $\cW(M, \g)$. That this 
map is a *-homomorphism can easily be seen from the formulas 
\eqref{wprd} and \eqref{conpr}, 
together with the relation $\omega'(x, y) = \omega(\chi(x), \chi(y))$. 
That $\iota_\chi$ is injective follows from the definition. 
\end{proof}

\medskip
\noindent
{\it Remarks:}
(1) If $\omega''$ is an arbitrary quasi-free Hadamard state on $(N, \g')$, 
then, in terms of the generators $W_n''(t)$ of $\cW(N, \g')$ in the
GNS representation of $\omega''$, we have 
\begin{eqnarray}
\iota_\chi(W_n''(t)) = \sum_{k}
W_{n - 2k} (\langle d^{\otimes k}_\chi, t \rangle \circ \chi^{-1}),
\end{eqnarray}
where $d_\chi(x, y)  = \omega(\chi(x), \chi(y)) - \omega''(x,y)$
and where $\langle d^{\otimes k}_\chi, t \rangle$ is given by 
Eq. \eqref{intdef}. 

(2)
We note that the identifications provided by the maps $\iota_\chi$ are 
consistent in the following sense. 
Let $\chi_{1,2}: M_1 \to M_2$ and $\chi_{2, 3}: M_2 \to M_3$ be 
causality preserving isometric embeddings and 
$\chi_{1, 3} = \chi_{2, 3} \circ \chi_{1, 2}$. 
Then the corresponding homomorphisms
satisfy (in the obvious notation)
$$\iota_{1,3} = \iota_{2, 3} \circ \iota_{1, 2}.$$ 
\begin{defn}
A quantum field $\Phi$ (in one variable) is an assignment which associates with every 
globally hyperbolic spacetime $(M, \g)$ a distribution $\Phi[\g]$
taking values in the algebra $\cW(M, \g)$, i.e., a continuous linear map 
$\Phi[\g]: \cD(M) \to \cW(M, \g)$. 
\end{defn}
Using the identifications provided by Lem. \ref{2.1}, we can 
now state what we mean by $\Phi$ being a ``local, covariant quantum field''. 
\begin{defn}
\label{def2.1}
A quantum field $\Phi$ (in one variable) is said to be 
{\it local and covariant}, if it satisfies the
following property: 
Let $\chi$ be an isometric embedding map from a 
spacetime $(N, \g')$ into another spacetime $(M, \g)$ 
(so that in fact $\g' = \chi^* \g$) which is causality 
preserving. Let 
$\iota_\chi: \cW(N, \g') \to \cW(M, \g)$ be the corresponding homomorphism, 
defined in Lem. \ref{2.1}. 
Then 
\begin{equation}
\iota_\chi(\Phi[\chi^* \g](f)) = \Phi[\g](f \circ \chi^{-1}) \quad 
\text{for all $f \in \cD(N)$.}
\end{equation}
Local fields in $n$ variables are defined in a similar manner. We will 
sometimes omit the explicit dependence of the fields on the metric. 
\end{defn}
\noindent
{\it Remarks:}
(1) 
The above type of algebraic formulation of the locality/covariance 
property was suggested to us by K. Fredenhagen \cite{owf}. It is closely related
to a formulation of ``locality'' previously given in \cite[pp. 89--91]{w1} for the stress energy operator. Antecedents to this idea can be found 
in \cite{w3} and \cite{bsk3}. 

(2)
It should be noted that the above definition involves actually two logically 
distinct requirements, namely (a) that the quantum field 
$\Phi[\g]$ under consideration 
be given by a diffeomorphism covariant expression, and (b) that it be 
{\it locally} constructed from the metric. The second requirement is incorporated 
in the possibility to consider isometries $\chi$ which map a spacetime $N$ into 
a portion of a ``larger'' spacetime $M$. This allows one to contemplate 
a situation in which ``the metric is varied outside 
some globally hyperbolic subset $N$ of a spacetime $M$''.  Note 
that the ``covariance'' axiom of Dimock \cite{dim} effectively 
corresponds to property (a), but since his axiom applies 
only to {\it global} isometries, it does not impose the 
requirement that the field depends only {\it locally} on the 
metric (property (b)). 

(3)
To illustrate our notion of local, covariant fields and to show that locality is in fact not 
a trivial requirement, we now display an example of a field which fails to be local. 
We consider, for every spacetime $(M, \g)$, the 
operator-valued distribution $\Phi[\g] = \;\lno \varphi^2 \rno_{\omega_{(M, \g)}}$, 
viewed now as an element of the abstract 
algebra $\cW(M, \g)$, where $\omega_{(M,\g)}$ is a quasi-free Hadamard state. 
We claim that the field $\Phi$ is not a local, covariant field, no matter 
how one assigns states $\omega_{(M, \g)}$ with globally hyperbolic 
spacetimes $(M, \g)$. The crucial observation needed to prove this is that the locality 
requirement, Def. \ref{def2.1}, would imply the following consistency relation
between the two-point functions of the given family of quasi-free Hadamard states: 
\begin{equation}
\label{cons}
\omega_{(M, \g)}(\chi(x), \chi(y)) = \omega_{(N, \g')}(x, y) \quad 
\text{$\forall (x, y) \in N \times N$,}
\end{equation}
whenever $\chi: N \to M$ is an isometric embedding map between two spacetimes
$(N, \g')$ and $(M, \g)$ (so that in fact $\g' = \chi^* \g$). 
To see that it is impossible to satisfy this constraint, consider the 
spacetimes $(M, \g)$ and $(M, \g')$ such that $\g \equiv \g'$ everywhere outside some 
region $O$ with compact closure. 
Let $\omega_{(M, \g)}$ and $\omega_{(M, \g')}$ be the quasi-free 
Hadamard states associated with those spacetimes. 
Let us now choose a Cauchy surface $\Sigma_+$ to the future of $O$ and a Cauchy surface 
$\Sigma_-$ to the past of $O$. Furthermore let 
us choose globally hyperbolic neighborhoods $N_\pm$ of $\Sigma_\pm$, which do 
not intersect $O$. 
The consistency requirement, Eq. \eqref{cons}, 
applied to the embeddings of $(N_\pm, \g)$ into the spacetimes 
$(M, \g)$ resp. $(M, \g')$ then immediately gives that
$\omega_{(M, \g)}(x, y) = \omega_{(N_\pm, \g)}(x, y)$ 
for all $(x, y) \in N_\pm \times N_\pm$ and that 
$\omega_{(M, \g')}(x, y) = \omega_{(N_\pm, \g)}(x, y)$
for all $(x, y) \in N_\pm \times N_\pm$. From this we get    
\begin{equation}
\omega_{(M, \g)}(x, y) = \omega_{(M, \g')}(x, y) \quad \text{$\forall 
(x, y) \in N_+ \times N_+$
and $\forall (x, y) \in N_- \times N_-$}.
\end{equation}  
This means that the two-point functions of the states $\omega_{(M, \g)}$ and 
$\omega_{(M, \g')}$ have the same 
initial data both on $\Sigma_+$ and $\Sigma_-$. But they do not obey the same field 
equation (the metrics $\g$ and $\g'$ being different inside $O$). From this 
one can easily obtain a contradiction. 

The above argument can be applied to 
any normal ordered operator, in particular to the normal ordered stress energy 
tensor. Our argument therefore gives a precise meaning to the common 
statement that normal ordering is not a valid procedure 
for defining the quantum stress-energy tensor in curved spacetime: 
The normal ordered stress tensor is not a local, covariant field. 

\medskip
\noindent
For later purposes, we also find it useful to make the following definition. 

\begin{defn}
\label{gammatdef}
Let $\Phi(x_1, \dots, x_n)$ be a local, covariant field in $n$ variables. Then, for 
any globally hyperbolic spacetime, $(M, \g)$, we define a conic subset $\Gamma^\Phi(M, \g)
\subset (T^*M)^n \backslash \{0\}$ associated with $\Phi$ by  
\begin{eqnarray}
\label{wunion}
\Gamma^\Phi(M, \g) \mydef \;\overline{\bigcup_{\omega} \WF(\omega(\Phi[\g]( \, \cdot \,)))}, 
\end{eqnarray}
where the closure is taken in $(T^*M)^n \backslash \{0\}$, and where the union 
runs over all quasi-free Hadamard states. 
\end{defn}
\noindent
{\it Remark:} If $\chi$ is a global diffeomorphism of $M$, then we have $\Gamma^\Phi(M, \chi^*\g)
= \chi^* \Gamma^\Phi(M, \g)$. This is a straightforward consequence of our notion of
local, covariant fields.

\section{Additional properties of local Wick polynomials and their time ordered products}
\label{sec3}

As we have seen, although normal ordering is mathematically a well defined prescription for 
defining powers of field operators, it does not define a local, covariant field, and is therefore 
not of any particular physical interest. Consequently, the same also applies to 
time ordered products of normal ordered Wick powers. In particular, 
the latter should not be used for the perturbative definition of an interacting field theory, 
since this field theory would then depend on nonlocal information, namely the global properties 
of the state chosen for the normal ordering prescription. We therefore seek to define 
a notion of {\it local} Wick polynomials and {\it local} time ordered products in the algebras $\cW(M, \g)$.   
In the present section, we shall specify these fields axiomatically (but not uniquely, as we 
shall see) by certain properties, which can heuristically be stated as follows: 

\medskip
\noindent
(i) {\bf Locality:} The sought-for Wick products and time ordered products are 
local, covariant fields in the sense of Def.~\ref{def2.1}.

\medskip
\noindent
(ii) {\bf Specific properties:} They have properties analogous to certain properties 
known to hold for the normal ordered Wick products and the time ordered products of these, 
such as for example a specific expression for their commutator with a free field.  

\medskip
\noindent
(iii) {\bf Continuity and Analyticity:} The fields vary analytically (continuously) 
under analytic (smooth) variations of the metric 
and the coupling parameters.

\medskip
\noindent 
(iv) {\bf Scaling:} The fields scale homogeneously ``up to logarithmic terms'' 
under a rescaling of the metric and the coupling parameters.

\vspace{0.5cm}

We have given a precise definition of requirement (i) in the previous section.
A mathematically precise formulation of conditions (ii)---(iv) will now be 
given in the following three subsections. 

\subsection{Specific properties}
\label{sec3.0}

We first consider local Wick powers of the free field without derivatives. 
These are denoted by $\varphi^k$, 
where $k \in \mn$. We make the obvious requirement that $\varphi^1$
be identical with the free field $\varphi$ (which is easily checked to be a local, 
covariant field), 
and for later convenience we also set
$\varphi^0 = 1$. We impose the following 
conditions on $\varphi^k$: 

\medskip
\noindent
{\bf Expansion:}
$[\varphi^k(x), \varphi(y)] = 
\i k\Delta_\g(x, y)\varphi^{k-1}(x)$. 

\medskip
\noindent
{\bf Hermiticity:} $\varphi^k(f)^* = \varphi^k(\bar f)$ for all $f \in \cD(M)$. 

\medskip
\noindent
{\bf Microlocal spectrum condition:} Let $\omega$ be a quasi-free Hadamard state. Then
$\omega(\varphi^k(x))$ is a smooth function in $x$. 

\medskip

Local Wick powers of differentiated fields are required to satisfy suitably 
generalized versions of the above requirements. The modifications are straightforward
and therefore left to the reader. For notational simplicity we will explicitly consider 
only the undifferentiated Wick powers in the following, but
our existence and uniqueness arguments and
results apply to the differentiated Wick powers as well as to the undifferentiated 
Wick powers. 

\noindent
{\it Remark}: For the 
local Wick products of differentiated fields it also would be reasonable to impose the 
following additional requirement: Any local Wick product containing $(\square - \xi R - m^2)\varphi$
as a factor should vanish. We note that the explicit construction of local 
Wick products that will be given in Sec. \ref{sec5.2} does not satisfy that requirement. 
(A related difficulty with our prescription 
given in Sec. \ref{sec5.2} is that it gives a stress energy operator which is 
not conserved.) 
We believe that a construction of local Wick products of differentiated 
fields satisfying this  additional condition can be given via 
the use of the local vacuum-concept introduced by Kay \cite{k} 
(see also \cite[Ch. 6]{hol}), but we will defer the consideration of this issue to 
a future investigation. 

\medskip

We next consider local time ordered products of undifferentiated local Wick powers. These 
are denoted by $T(\varphi^{k_1} \dots \varphi^{k_n})$. We make the 
obvious requirement that $T(\varphi^k)$ be equal to the
local Wick power $\varphi^k$ considered above. Our further requirements
are the following

\medskip
\noindent
{\bf Symmetry:}
Any time ordered product is symmetric under a permutation of the operators 
under the time-ordering symbol. 

\medskip
\noindent
{\bf Causal factorization:}
Consider any set of points $(x_1, \dots, x_n) \in M^n$ and a partition of $\{1, \dots, n\}$
into two non-empty subsets $I$ and $I^c$, with the property that no point $x_i$ with
$i \in I$ is in the past of any of the points $x_j$ with $j \in I^c$, i.e., 
$x_i \notin J^-(x_j)$ for all $i \in I$ and  $j \in I^c$. Then the time ordered products 
factorize in the following sense:
\begin{eqnarray*}
T(\varphi^{k_1}(x_1) \dots \varphi^{k_n}(x_n))                         
= T\left(\prod_{i\in I} \varphi^{k_i}(x_i)\right)\,T\left(\prod_{j \in I^c} \varphi^{k_j}(x_j)\right).
\end{eqnarray*}

\medskip
\noindent
{\bf Expansion:}
\( \displaystyle{
[T(\varphi^{k_1}(x_1) \dots \varphi^{k_n}(x_n)), \varphi(y)] =}\)
\begin{flushright}
\(\displaystyle{
\i\sum_{i=1}^n k_i \Delta_\g(x_i, y)
T( \varphi^{k_1}(x_1)\dots 
\varphi^{k_i-1}(x_i)  \dots \varphi^{k_n}(x_n)).}  
\)
\end{flushright}

\medskip
\noindent
{\bf Unitarity:}
\(\displaystyle{
T(\varphi^{k_1}(x_1) \dots \varphi^{k_n}(x_n))^* = \sum_{\cP = I_1 \uplus \dots \uplus I_j}
(-1)^{n + j} \prod_{I \in \cP} T\left( \prod_{i \in I} \varphi^{k_i}(x_i)\right)}\).

\noindent
Here we have used the following notation: $\cP = I_1 \uplus \dots \uplus I_j$ denotes a partition of 
the set $\{1, \dots, n\}$ into $j$ pairwise
disjoint, nonempty subsets $I_i$. 
The unitarity condition is equivalent to requiring that the $S$-matrix is unitary in 
the sense of formal power series of operators.

\medskip
\noindent
{\bf Microlocal spectrum condition:}
Let $\Gamma^T(M, \g) \subset (T^*M)^n \backslash \{0\}$ 
be the conic set associated with the time ordered product
$T(\varphi^{k_1}(x_1) \dots \varphi^{k_n}(x_n))$ as in Def. \ref{gammatdef}.  Then, any 
point $(x_1, k_1, \dots, x_n, k_n)$ 
in $\Gamma^{T}(M, \g)$ satisfies the following: (a) 
there exist null-geodesics $\gamma_1, \dots, \gamma_m$ which 
connect any point $x_j$ in the set $\{x_1, \dots, x_n\}$ to 
some other point in that set, 
(b) there exists coparallel, cotangent covectorfields $p_1, 
\dots, p_m$ along these geodesics such that $p_i \in V^+$
if the starting point of $\gamma_i$ is not in the causal 
past of the end point of $\gamma_i$, (c) for the covector 
$k_j$ over the point $x_j$ it holds that 
$k_j = \sum_e p_e(x_j) - \sum_s p_s(x_j)$, where the index
$e$ runs through all null-geodesics ending at $x_j$ and $s$ runs
through all null-geodesics starting at $x_j$. 

The microlocal spectrum condition may be viewed as a microlocal analogue of 
translation invariance in Minkowski space. It was shown to hold for time ordered
products of normal ordered Wick powers in \cite{bf}. We also note that it 
reduces to the requirement that $\omega(\varphi^k(x))$ be smooth in the case $n=1$. 

\medskip

Again, time ordered products of differentiated Wick powers would satisfy suitable 
generalizations of the above requirements. Our uniqueness arguments of Sec. \ref{mainpar}
would also apply to such time ordered products, but for notational simplicity 
we shall explicitly only consider the undifferentiated products below. 

\medskip 
For later purposes, we also wish to impose a sharpened version of the microlocal 
spectrum condition for the local Wick polynomials and their time ordered products
for the case that the metric $\g$ is not only smooth, but 
in addition real analytic in some convex normal neighborhood $O \subset M$. For 
this purpose, we consider ``analytic'' quasi-free 
Hadamard states, i.e., quasi-free states 
$\omega$ with the property that $\omega(x, y) - H(x, y)$ 
is not only a smooth, but in addition an analytic function in $O \times O$, 
where $H$ is the Hadamard fundamental solution defined by Eq. \eqref{Hdef}. 
We then impose a sharpened constraint on the singular behavior of the expectation 
values of a local time ordered product in such a state by considering the so-called 
``analytic wave front set'' \cite{h} instead of the 
ordinary, ``smooth wave front set'', which is used in the above microlocal 
spectrum condition (compare Def. \ref{gammatdef}). The concept of the 
analytic wave front set, $\WF_A(u)$, of a distribution $u$ characterizes 
the points and directions for which $u$ fails to be analytic, in much the 
same way as the ordinary wave front set, $\WF(u)$, characterizes the points 
and directions for which $u$ is not smooth.\footnote{We note that 
for any distribution $u$ it holds that $\WF(u) \subset \WF_A(u)$.}  

In order to give a formulation of the microlocal spectrum condition in the analytic case 
that is parallel to the one given above in the smooth case, we first introduce, 
for every local, covariant field $\Phi(x_1, \dots, x_n)$, a conic set $\Gamma^\Phi_A(O, \g)
\subset (T^*O)^n \backslash \{0\}$, which 
is defined as in Def. \ref{gammatdef}, but with the difference that the 
union in Eq. \eqref{wunion} now runs over all analytic Hadamard states in $O \times O$, 
and that $\WF$ is replaced by $\WF_A$. In the case when $\Phi(x_1, \dots, x_n)$ is 
a local time ordered product, we denote this conic set by $\Gamma^T_A(M, \g)$. 
Our analytic microlocal spectrum condition is 
then the following: 

\medskip
\noindent
{\bf Analytic microlocal spectrum condition:} Let $O$ be a convex normal neighborhood 
of $M$. Then any point $(x_1, k_1, \dots, x_n, k_n) \in \Gamma^T_A(O, \g)$ has
the properties stated in the microlocal spectrum condition for the smooth case.  

\medskip
\noindent
{\it Remark:} For a local Wick product (the case $n=1$), this condition implies that $\omega(\varphi^k(x))$ is 
analytic in $O$ for any analytic Hadamard state.

\subsection{Continuity and analyticity}
\label{sec3.1}

The basic difficulty in defining notions of continuous and analytic dependence of a local, covariant 
field under a corresponding variation of the metric and the parameters is that the 
fields corresponding to different metrics and parameters are elements of 
different algebras and hence cannot be compared directly. 
It is therefore necessary to provide a suitable identification of these elements first. 
In order to simplify the discussion, we will first consider only variations of the 
spacetime metric, and keep the coupling constants fixed. We will comment on how to 
generalize the present discussion to include also variations of the parameters at the 
end of this subsection. 

We first give a notion of the continuous dependence of a local, covariant quantum
field on the metric. Here, we consider a situation wherein one is given a family of metrics, $\g^{(s)}$, 
depending smoothly on some real parameter, $s$, and differing from each 
other only within some compact region, 
$O$, in the spacetime $M$. 
Under these circumstances, we will show in Lem. \ref{3.1} that it is possible to construct 
isomorphisms between the algebras corresponding to different values of $s$  
by identifying the observables in the past (or future) of $O$. A local, covariant 
field $\Phi$ with a continuous dependence under smooth variations of the 
metric will then be defined as one for which the family $\Phi[\g^{(s)}]$ 
depends continuously on $s$ under this identification of the corresponding
algebras for all smooth families of metrics $\g^{(s)}$. 

A notion of the analytic dependence of a local, covariant field under corresponding variations of
the metric is given next. Here, we consider an analytic family, $\g^{(s)}$, 
of real analytic metrics in some open neighborhood $O$ of $M$. 
However, unlike in the case of a {\it smooth} family of metrics considered above, 
we now cannot demand that our metrics coincide outside some
compact region, because there are no analytic functions with compact support. 
Consequently, we cannot identify the algebras for different values of $s$ in 
the same manner as in the smooth case, and we therefore have no obvious means to compare directly a given 
field for the different metrics $\g^{(s)}$, since these fields belong to different algebras.
We will avoid this problem by considering instead a notion of analytic dependence of a field on
the metric via its expectation values in an analytic family of quasi-free Hadamard states, 
$\omega^{(s)}$, corresponding to the metrics $\g^{(s)}$: We shall say that 
a local, covariant field $\Phi$ depends analytically on the metric if the family of expectation values
$\omega^{(s)}(\Phi[\g^{(s)}](x_1, \dots, x_n))$ depends, in a suitable sense, analytically 
on $s$, for all possible choices of analytic families of metrics $\g^{(s)}$ and 
states $\omega^{(s)}$. 

\begin{lemma}
\label{3.1}
Consider two globally hyperbolic spacetimes $(M, \g)$ and $(M, \g')$, 
such that $\g \equiv \g'$ everywhere outside some region $O$ with compact closure. 
Then there exists a *-isomorphism 
$\tau_{\rm ret}: \cW(M, \g') \to \cW(M, \g)$, such that the 
restriction of $\tau_{\rm ret}$ to the subalgebra $\cW(M_-, \g')$ with 
$M_- = M \backslash J^+(O)$ is the identity.
Similarly there exists a *-isomorphism 
$\tau_{\rm adv}: \cW(M, \g') \to \cW(M, \g)$, such that the 
restriction of $\tau_{\rm adv}$ to the subalgebra $\cW(M_+, \g')$ 
with $M_+ = M \backslash J^-(O)$ is the identity.
\end{lemma}
\noindent
{\it Remark:}
The isomorphisms $\tau_{\rm ret}$ and $\tau_{\rm adv}$ are constructed 
by a suitable identification of the fields in both algebras on a Cauchy surface 
$\Sigma_-$ not intersecting the future of $O$ or, respectively, on a 
Cauchy surface $\Sigma_+$ not intersecting the past of $O$. 
The particular choice of those Cauchy surfaces 
is irrelevant for the constructions, so in that sense, $\tau_{\rm ret}$ and 
$\tau_{\rm adv}$ are
canonical. In the following proof, we will only construct 
$\tau_{\rm ret}$, the construction of $\tau_{\rm adv}$ is completely 
analogous.
\begin{proof}
Let $\Sigma_-$ be a Cauchy surface not intersecting the future of 
$O$ and let $\Sigma_+$ be a Cauchy surface not intersecting the past of $O$.
Define a bidistribution $S$ on $M$ by 
\begin{eqnarray}
S(f_1 \otimes f_2) = \int_{\Sigma_-} (F_1 {\nabla}_a F_2 - F_2 \nabla_a F_1)\,n^a d\sigma, 
\end{eqnarray}
where 
\begin{eqnarray}
F_1(x) = \int_M \Delta_{\g}(x, y) f_1(y) \, \mu_{\g}(y), \quad 
F_2(x) = \int_M \Delta_{\g'}(x, y) f_2(y) \, \mu_{\g'}(y).
\end{eqnarray}
By a standard argument based on  Gauss' law (see e.g. \cite{w1}), one can see that 
$S$ does not depend on the particular choice for $\Sigma_-$.
Let $\chi$ be an arbitrary smooth function on $M$ satisfying 
$\chi(x) = 0$ for all $x \in J^+(\Sigma_+)$ and  
$\chi(x) = 1$ for all $x \in J^-(\Sigma_-)$.
We then define a linear map $A_{\rm ret}: \cD(M) \to \cD'(M)$ by 
$$A_{\rm ret}f  \mydef -(\square_\g - \xi R_\g - m^2)(\chi Sf).$$ 
The distribution $A_{\rm ret}f$ satisfies the following properties: 
\begin{enumerate}
\item[(a)]
$A_{\rm ret}f$ is of compact support with ${\rm supp}(A_{\rm ret}f) \subset J^+(\Sigma_-) \cap J^-(\Sigma_+)$, 
\item[(b)]
$\Delta_{\g} A_{\rm ret}f(x) = \Delta_{\g'} f(x)$ for all $x \in J^-(\Sigma_-)$ and $f \in \cD(M)$.
\end{enumerate}
Item (a) immediately follows from the fact that $(\square_\g - \xi R_\g - m^2)Sf(x) = 
0$ for all $x \in J^-(\Sigma_-)$ and the fact that $\chi(x) = 0$ for all 
$x \in J^+(\Sigma_+)$. Item (b) holds since   
\begin{equation}
\Delta_{\g} A_{\rm ret}f(x) = 
\Delta_{\g}^{{\rm ret}} (\square_\g - \xi R_\g - m^2)(\chi S f)(x) =
Sf(x) = \Delta_{\g'} f(x) \quad \forall x \in J^-(\Sigma_-).
\end{equation}

We wish to show that the 
$n$-th tensor power of $A_{\rm ret}$ gives a map $$A_{\rm ret}^{\otimes n}: 
\cE_n'(M,\g') \to \cE'_n(M, \g).$$  
We begin by showing that $S$ has the following wave front set: 
\begin{eqnarray}
\WF(S) &\subset& 
\{ (x_1, k_1, x_2, -k_2) \in (T^*M)^2 \backslash \{0\} 
\mid 
\text{$\exists y \in M \backslash J^+(O)$ 
and $(y, p) \in T^*_y M$} \nonumber \\
&&\text{such that $(x_1, k_1) \sim (y, p)$ with respect to  $\g$ and such that}\nonumber\\ 
&&\text{$(x_2, k_2) \sim (y, p)$ with respect to $\g'$} \}.
\end{eqnarray}
In order to see this, we note that by definition, 
\begin{equation}
(\square_\g - \xi R_\g - m^2)_x S(x, y) = 
(\square_{\g'} - \xi R_{\g'} - m^2)_y S(x, y) = 0.
\end{equation}
We are thus in a position to apply the ``propagation of singularities theorem'' 
\cite[Thm. 6.1.1]{h1} to $S$. 
This theorem tells us that  
an element $(x_1, k_1, x_2, k_2)$ is in $\WF(S)$ if and only if every element of the 
form $(y_1, p_1, y_2, p_2)$ is in $\WF(S)$, where $(y_1, p_1) \sim (x_1, k_1)$ with 
respect to $\g$ and where $(y_2, p_2) \sim (x_2, k_2)$ with 
respect to $\g'$. Moreover, by definition of $S$, 
we have that $S(x, y) = \Delta_{\g}(x, y) = \Delta_{\g'}(x, y)$
for all $x, y \in M \backslash J^+(O)$. 
The wave front set of $\Delta_\g$ is known to be 
\begin{eqnarray}
\WF(\Delta_\g) = 
\{ (x_1, k_1, x_2, -k_2) \in (T^*M)^2 \backslash \{0\} 
\mid \text{$(x_1, k_1) \sim (x_2, k_2)$ with respect to $\g$}\}. 
\end{eqnarray}
Combining these two pieces of information then gives us the above  wave front set for 
$S$. 

Since differentiating and multiplying a distribution  
by a smooth function does not enlarge its wave front set, it holds that
$\WF(A_{\rm ret}) \subset \WF(S)$. By the rules \cite{h} 
for calculating the wave front set of a tensor product of distributions, we get
from this that
\begin{multline}
\WF(A^{\otimes n}_{\rm ret}) 
\subset 
\{(x_1, k_1, \dots, x_n, k_n, y_1, p_1, \dots, y_n, p_n) \in 
(T^*M)^{2n} \backslash \{0\} \mid \\
\text{$(x_i, k_i, y_i, p_i) \in \WF(S) \cup \{0\}$ for all 
$i = 1, \dots, n$}\}.  
\end{multline} 
Let $t \in \cE'_n(M, \g')$, that is, $t$ is a symmetric, compactly supported
$n$-point distribution with $\WF(t) \subset \G_n(M, \g')$. 
Then it follows from the above form of $\WF(A_{\rm ret}^{\otimes n})$ that 
\begin{multline}
\{(y_1, p_1, \dots, y_n, p_n) \in 
(T^*M)^n \backslash \{0\} \mid \\
\text{$\exists (x_1, 0, \dots, x_n, 0, y_1, -p_1, \dots, y_n, -p_n) \in \WF(A_{\rm ret}^{\otimes n})$}\}  
\cap \WF(t)
= \emptyset.
\end{multline}
Therefore \cite[Thm. 8.2.13]{h} applies and we conclude from that theorem that the    
linear operator $A_{\rm ret}^{\otimes n}$ has a well-defined action 
on distributions $t \in \cE'_n(M, \g')$. The wave front set 
of the distribution $A_{\rm ret}^{\otimes n}t$ can be calculated from \cite[Thm. 8.2.13]{h}
using our knowledge about $\WF(A_{\rm ret}^{\otimes n})$ and $\WF(t)$:
\begin{eqnarray}
\WF(A_{\rm ret}^{\otimes n} t) &\subset&
\{ (x_1, k_1, \dots, x_n, k_n) \in (T^*M)^{n} \backslash \{0\}
\mid \text{$\exists (x_i, k_i, y_i, -p_i) \in \WF(S) \cup \{0\}$,} 
\nonumber\\
&& \text{$i = 1, \dots, n,$ such that   
$(y_1, p_1, \dots, y_n, p_n) \in \G_n(M, \g')$} \} \nonumber\\
&\cup&
\{(x_1, k_1, \dots, x_n, k_n) \in (T^*M)^{n} \backslash \{0\} \mid 
\exists (x_i, k_i, y_i, 0) \in \WF(S) \cup \{0\} \nonumber\\
&&\text{for all $i =1, \dots, n$}
\} \nonumber\\
&\subset& \G_n(M, \g).  
\end{eqnarray}
Since the distribution $A_{\rm ret}^{\otimes n}t$ is of compact support by (a), we have 
thus demonstrated that the $n$-th tensor power of $A_{\rm ret}$ gives a map 
from $\cE_n'(M, \g')$ to $\cE'_n(M, \g)$, as we had claimed.  

The algebras $\cW(M, \g)$ and $\cW(M, \g')$ 
are faithfully represented on the GNS Hilbert spaces of any quasi-free Hadamard states $\omega$ 
respectively $\omega'$ on the 
corresponding Weyl subalgebras. We may choose 
these quasi-free states (or rather their two-point functions) to 
have identical initial data on $\Sigma_-$. In view of item (b), this amounts to 
saying that
\begin{eqnarray}
\label{wwtld}
\omega(A_{\rm ret}f_1 \otimes A_{\rm ret}f_2) = 
\omega'(f_1 \otimes f_2)
\end{eqnarray}
for all compactly supported test functions $f_1, f_2$. We now define 
$\tau_{\rm ret}: \cW(M, \g') \to \cW(M, \g)$ by 
\begin{eqnarray}
\tau_{\rm ret}(W_n'(t)) \mydef W_n(A_{\rm ret}^{\otimes n} t),  
\end{eqnarray}
where the $W_n'$ are the generators of $\cW(M, \g')$ and where the $W_n$
are the generators of $\cW(M, \g)$. 
We must show that this is indeed a *-isomorphism. That $A_{\rm ret}$ respects 
the product in both algebras, Eq. \eqref{wprd}, follows from 
\begin{equation}
A_{\rm ret}^{\otimes (n + m - 2k)}( t \otimes_k' t') = 
(A_{\rm ret}^{\otimes m}t) \otimes_k^{\,} (A_{\rm ret}^{\otimes n}t'), 
\end{equation}
where $\omega'$ is used for the contractions in 
$\otimes_k'$ on the left side, and $\omega$ is used for the contractions in $\otimes_k$ on 
the right side, as one can easily verify using relation 
Eq. \eqref{wwtld} and the definition of the contracted tensor product. That 
$\tau_{\rm ret}$ respects the *-operation follows because $A_{\rm ret}$ is 
real. That $\tau_{\rm ret}$ is invertible can be seen by an explicit construction 
of its inverse, given by the same construction as above, but with the 
spacetimes $(M, \g)$ and $(M, \g')$ interchanged. 
The definition of $A_{\rm ret}$ does not depend on the specific choice for 
$\Sigma_-$, but it depends on a choice for $\chi$. 
It is however not difficult to see that isomorphism $\tau_{\rm ret}$ itself is independent of that 
choice. We finally prove that the restriction of $\tau_{\rm ret}$ to $\cW(M_-, \g')$ is the 
identity. By item (b) above we have 
\begin{equation}
\Delta_\g(t - A_{\rm ret} t) = \Delta_\g t - \Delta_{\g'} t  \quad
\text{in $J^-(\Sigma_-)$} 
\end{equation}
for any $t \in \cE'_1(M, \g')$. Now if the support of $t$ is in $M_-$ (so that 
$\supp(t) \cap J^+(O) = \emptyset$)  then the above expression vanishes on $J^-(\Sigma_-)$. 
Since this expression is moreover a solution to the Klein-Gordon equation, it must in fact vanish everywhere. 
Therefore, by the same argument as in the proof of Prop.~\ref{pro1}, there is  
an $s \in \cE'_1(M, \g)$ such that $t - A_{\rm ret}t = (\square_\g - \xi R_\g - m^2) s$. 
Since $W_1((\square_\g - \xi R_\g - m^2) s)=0$, this implies that 
$\tau_{\rm ret}(W_1'(t)) = W_1(A_{\rm ret} t) = W_1(t)$ for all $t \in \cE'_1(M_-, \g')$. 
This argument can be generalized to show that 
$\tau_{\rm ret}(W_n'(t)) = W_n(t)$ for all $t \in \cE'_n(M_-, \g')$ and 
arbitrary $n$, 
thus proving our claim. 
\end{proof}

Using the above lemma, we are now able to say what precisely we mean by the statement that 
a ``local field varies continuously under a smooth variation 
of the metric''. Let $\g^{(s)}$ be a family of metrics on $M$ such that 
$\g^{(s)} \equiv \g$ outside a compact region $O$ and which depends smoothly
on $s$ in the sense that the five-dimensional metric $g_{ab}^{(s)} + (ds)^{\,}_a (ds)^{\,}_b$ 
is smooth on $M \times \mr$. From the above lemma, we then get, for each value of $s$, an 
isomorphism $\tau_{\rm ret}: \cW(M, \g^{(s)}) \to \cW(M, \g)$.

\medskip
\noindent
{\bf Continuity:}
A local, covariant quantum field $\Phi$ is said to depend continuously on 
the metric if the algebra-valued function 
$$
\mr \owns s \to \tau_{\rm ret}\Big(\Phi[\g^{(s)}](f)\Big) \in \cW(M, \g)
$$
is continuous for all families
of metrics as described above and all test functions $f$.

\medskip
\noindent
{\it Remarks}: (1)
A notion of continuous dependence of the fields on the metric could also be 
given based on the isomorphisms $\tau_{\rm adv}$. It can be seen (although we 
do not demonstrate this here) that both notions coincide.

(2) We also note that the isomorphisms $\tau_{\rm adv}$ and $\tau_{\rm ret}$ can be used in certain 
cases to describe in 
a meaningful way the advanced and retarded 
response of local, covariant quantum field to an infinitesimal perturbation of 
the metric in the past and future. Namely, for a local, covariant field 
$\Phi$ which has not only a continuous but in addition 
a once differentiable dependence on the metric, one can define 
its advanced response, $(\delta\Phi/\delta g_{ab})_{\rm adv}$, 
to a metric perturbation by 
\begin{multline}
\int_{M^{n+1}}
\left(\frac{\delta \Phi(x_1, \dots, x_n)}{\delta g_{ab}(y)}
\right)_{\rm adv}{h}_{ab}(y) f(x_1, \dots, x_n) \; \mu_\g(y)\mu_\g(x_1) \dots
\mu_\g(x_n)\\
\mydef\; 
\frac{d}{ds} \tau_{\rm adv}(\Phi[\g + s {\bf h}](f)) \Big|_{s=0}, 
\end{multline}
where ${\bf h} \equiv h_{ab}$ is of compact support.
In the same way one can define the retarded response, 
$(\delta\Phi/\delta g_{ab})_{\rm ret}$,  
of a local, covariant field $\Phi$ to a metric perturbation.

\medskip
\noindent
We next explain what we mean by the statement that a ``local field varies analytically under an
analytic variation of the metric''. Let $\g^{(s)}$ be a family of metrics on $M$ which 
is analytic in some convex normal neighborhood $O \subset M$ in the sense 
that the five-dimensional metric $g_{ab}^{(s)} + (ds)^{\,}_a (ds)^{\,}_b$ is analytic
on $O \times (-\epsilon, \epsilon)$. We consider a family of quasi-free Hadamard
states, $\omega^{(s)}$, on the algebras $\cW(M, \g^{(s)})$ that is 
analytic in $s$ in the following sense: Let $H^{(s)}$ be the Hadamard
parametrices, given by Eq. \eqref{Hdef}, constructed from the metrics $\g^{(s)}$, and 
let us assume that $O$ is small enough such that $H^{(s)}$ is well-defined
on $O \times O$ for all $s$. We say that $\omega^{(s)}$ is an analytic 
one-parameter family of states if the difference 
$\omega^{(s)}(x, y) - H^{(s)}(x, y)$ is
jointly analytic in $(x, y, s)$ on $O \times O \times (-\epsilon, \epsilon)$. 
We would like to define a notion of the analytic dependence of a local 
field on the metric by demanding that the  
expectation values $\omega^{(s)}(\Phi[\g^{(s)}](x_1, \dots, x_n))$ depend 
analytically on $s$ for any analytic family of metrics and any corresponding 
analytic family of quasi-free Hadamard states. However, since these expectation values are in fact 
distributions in $x_1, \dots, x_n$, it is 
not clear a priori what is actually meant by ``analytic dependence on $s$''. To 
give precise meaning to this statement we must characterize the 
extent to which the above expectation values, viewed as distributions
jointly in $(x_1, \dots, x_n, s)$, ``fail to be analytic''. 
We do so by means of the analytic 
wave front set of the above expectation values of a local, covariant field, viewed
as a distribution jointly in $(x_1, \dots, x_n, s)$. 

\medskip
\noindent
{\bf Analytic dependence:} Let $\g^{(s)}$ be an analytic family of metrics in 
$O \subset M$ and let $\omega^{(s)}$ be a corresponding analytic family of 
quasi-free Hadamard states. Let $\Phi$ be a local, covariant field in 
$n$ variables, and let $\Gamma^\Phi_A(O, \g) \subset (T^*O)^n \backslash \{0\}$ be 
the associated conic set as introduced in Subsection \ref{sec3.0}. Consider 
the family of expectation values, 
\begin{equation}
E_\omega^\Phi(x_1, \dots, x_n, s) \mydef \omega^{(s)}
\Big(\Phi[\g^{(s)}](x_1, \dots, x_n)\Big),
\end{equation} 
viewed as a distribution on $O^n \times (-\epsilon, \epsilon)$. 
Then we demand that 
\begin{multline}
\WF_A(E^\Phi_\omega) \subset \{ 
(x_1, k_1, \dots, x_n, k_n, s, \rho) \in T^*(O^n \times (-\epsilon, 
\epsilon))
\mid \\ 
(x_1, k_1, \dots, x_n, k_n) \in \Gamma^\Phi_A(O, \g^{(s)})\} 
\end{multline} 
for all analytic families of metrics and all corresponding analytic 
families of states. 

\medskip
\noindent
{\it Remarks}: (1) The above condition on the analytic wave front set can 
be understood as follows. Consider first an open neighborhood 
$U \subset O^n$ such 
that $E^\Phi_\omega$ is non-singular for all $(x_1, \dots, x_n) \in U$
for a given value of $s = s_0$. Then the condition on $\WF_A(E^\Phi_\omega)$ implies that 
$E^\Phi_\omega$ varies analytically in $(x_1, \dots, x_n)$ and $s$ in neighborhood of the form 
$U \times (s_0-\delta, s_0+\delta)$ for some $\delta > 0$. On the other hand, 
if $(x_1, \dots, x_n)$ is a singular point for the local, covariant field 
$\Phi$
at a given $s$, then the condition on $\WF_A(E^\Phi_\omega)$ demands that the 
singular ``$x$-directions'' of $E^\Phi_\omega$ in momentum space are the same
ones as for the field $\Phi[\g^{(s)}](x_1, \dots, x_n)$, considered as 
a distribution in the $x$-variables at fixed $s$. 

(2) The above definition assumes the existence of an analytic family of states for 
any given analytic family of metrics. While we do not have any argument 
proving the existence of such a family, we remark that, for the sake of our definition 
of analytic dependence, it would be entirely sufficient to have a 
suitable family, $\psi^{(s)}$, of normalized, 
linear (but not necessarily positive) functionals on the algebras 
$\cW(M, \g^{(s)})$. We now briefly indicate how such a family can be 
constructed. Firstly, using the results of \cite[Ch. 6]{hol} one can 
obtain families of bidistributions
$\psi^{(s)}(x, y)$ which have the same properties as $\omega^{(s)}(x, y)$, 
except possibly for positivity. These bidistributions can then be promoted, by the 
same formula as Eq. \eqref{npoint}, to normalized linear functionals on the 
algebras $\cA(M, \g^{(s)})$ of free fields. It is then not 
difficult to see that these 
can then be extended (via normal ordering elements of $\cW(M, \g^{(s)})$ with 
respect to $\psi^{(s)}$) to functionals on the algebras $\cW(M, \g^{(s)})$. 

\medskip

The analyticity of local, covariant fields under corresponding variations of the 
coupling parameters can be formulated in a very similar way as above. To obtain 
a corresponding notion of continuous dependence, it is however 
necessary to allow the coupling parameters $p$ ($ \equiv (\xi, m^2)$ in the 
case of a real scalar field, Eq.~\eqref{action})
to be arbitrary smooth functions on spacetime, rather than constants. One can then 
consider two coupling functions $p_1$ and $p_2$ which differ only within some 
compact region. In such a situation, it is possible to find an identification of 
the algebras corresponding to $p_1$ and $p_2$, which is analogous to the one established in
Lem. \ref{3.1}. Based on such an identification, one can give a 
notion of continuity of local, covariant fields 
under smooth variations of the coupling parameters, which is completely analogous 
to the above notion of continuity under smooth variations of the metric. 
It should also be noted that the  consideration of different coupling parameters involves 
a slight generalization of our notion of local, covariant fields (Def. \ref{def2.1}). 
This generalization is however rather obvious and therefore left to the reader. 

\subsection{Scaling}
\label{sec3.2}
The scaling requirement involves the comparison of a given local, covariant field at different scales,
i.e., its behavior under a rescaling $\g \to \lambda^{-2} \g$ and under corresponding rescalings 
of the coupling parameters $m^2, \xi$ and $\varphi$, chosen in such a way as 
to leave the action $\cal S$ invariant. For the action \eqref{action}, the unique 
corresponding scalings of $m^2, \xi$ and $\varphi$ leaving $\cal S$ invariant are  
$m^2 \to \lambda^2 m^2, \xi \to \xi$ and $\varphi \to \lambda \varphi$. 
We will refer to the various exponents of $\lambda$ as the the ``engineering dimension'' 
of the corresponding quantities (and similarly for other quantities derived from those). 
In order to compare an arbitrary local, covariant field $\Phi$ in the algebras $\cW(M, \g)$ at different
scales, we first show that the algebras constructed from the rescaled quantities 
are naturally isomorphic for all values of $\lambda > 0$. 

\begin{lemma}
\label{3.2}
There are natural *-isomorphisms $\sigma_\lambda: 
\cW_{p(\lambda)}(M, \lambda^{-2}\g) \to \cW_p(M, \g)$ for all 
$\lambda > 0$,  
where the subscripts on the algebras indicate 
the dependence on the parameters, $p = (\xi, m^2)$ and $p(\lambda) = 
(\xi, \lambda^2 m^2)$.  
\end{lemma}
\begin{proof}
Let $\omega$ be a quasi-free Hadamard state for the theory at $\lambda = 1$. For all 
$\lambda > 0$, let
\begin{eqnarray}
\label{wla}
\omega^{(\lambda)}(x, y) = \lambda^{2}\omega (x, y).
\end{eqnarray}
Then $\omega^{(\lambda)}$ is the two-point function 
of a quasi-free Hadamard state
of the theory scaled by $\lambda$. (Note that Eq. \eqref{wla} 
is equivalent to the relation
$\omega^{(\lambda)}(f_1 \otimes f_2) = \lambda^{-6} \omega(f_1 \otimes
f_2)$ between the smeared two-point functions, because the metric
volume element transforms as $\mu_{\lambda^{-2}\g} = 
\lambda^{-4} \mu_{\g}$.)
We use $\omega^{(\lambda)}$ to give 
a concrete realization of the algebra $\cW_{p(\lambda)}(M, \lambda^{-2}\g)$. 
We then define 
(using the same symbol for the generators $W_n$ in both algebras)
\begin{eqnarray*}
\sigma_\lambda: \cW_{p(\lambda)}(M, \lambda^{-2}\g) \owns W_n(t) 
\to \lambda^{-3n} W_n(t) \in \cW_{p}(M, \g). 
\end{eqnarray*}
$\sigma_\lambda$ is a well defined map for all $\lambda > 0$, because $\cE'_n(M, \g) 
= \cE'_n(M, \lambda^{-2}\g)$. 
Using Eq. \eqref{wla}, it is also easily checked to be a *-homomorphism. 
\end{proof}
Using the above lemma, we are now in a position to consider a given local, covariant field at 
different scales: Let $\Phi$ be a local, covariant field in $n$ 
variables. We 
then define a rescaled field, $S_\lambda \Phi$, by
\begin{eqnarray}
\label{Slambda}
S_\lambda \Phi[\g, p](f) \mydef 
\lambda^{4n}\sigma_\lambda\Big( \Phi[\lambda^{-2} \g, p(\lambda)](f)\Big),
\end{eqnarray}
where $p(\lambda) = (\xi, \lambda^2 m^2)$ and $\lambda > 0$. 
The crucial point to note about the automorphism $\sigma_\lambda$ is that (a) it ensures 
that the field $\Phi$ and the rescaled field $S_\lambda\Phi$ live in 
the same algebra (so that they may be compared), and that (b) it is constructed in such a way 
that the rescaled field $S_\lambda \Phi$ is again local in the sense of Def. \ref{def2.1}. The factor $\lambda^{4n}$ has been included 
in the definition of the scaling map $S_\lambda$ in order 
to compensate for the fact that the quantum fields 
are distributions and therefore transform as 
densities under rescalings of the metric. 
The action of $S_\lambda$ on some simple 
local, covariant fields is given below. 

Next, we introduce the notion of the scaling dimension of a local, covariant field. 
\begin{defn}
The scaling dimension $d_\Phi$ of a local, covariant field $\Phi$ is defined by 
\begin{eqnarray}
d_\Phi = \inf \{\delta \in \mr \mid \lim_{\lambda \to 0+} 
\lambda^{-\delta} S_\lambda \Phi = 0 \}, 
\end{eqnarray}
where the limit is understood to mean that 
\begin{eqnarray*}
\lim_{\lambda \to 0+} \lambda^{-\delta} S_\lambda \Phi[\g, p](f) = 0 
\end{eqnarray*}
for all metrics $\g$, all values of the 
parameters $p$ and  all test functions $f$. 
\end{defn}
It is easy to see from the definition that the free field indeed scales as 
$S_\lambda \varphi = \lambda \varphi$. 
The local c-number field $C = m^2 R1$ scales as $S_\lambda C = \lambda^4 C$, 
so it has scaling dimension four. The fields in the above examples scale homogeneously. 
However, this is clearly not always so, as may be seen from the 
elementary example $(1 + R^2)^{-1}1$, which is local, has scaling dimension zero, but which does not 
scale homogeneously (and which also has no well-defined engineering dimension). 

We would like to require that our local Wick powers and local time ordered 
products scale homogeneously, the basic idea being that we 
wish our fields to have a well-defined engineering dimension. However, as it 
is well known in quantum field theory---and, as we shall see in more detail for the local Wick products
below---logarithmic terms cannot be avoided in general (with the exception of the free field).
Consequently, we will require, instead, that the local Wick powers and their local time ordered
products scale ``homogeneously up to logarithmic terms''. 
This requirement is formulated precisely as follows.
We say that an element $a \in \cW_{p}(M, \g)$ has 
{\it order} $k$ if its $(k+1)$ times repeated commutator with a free field vanishes. 
(Prop. \ref{pro1} provides a characterization of such elements.) 
By the expansion requirement, we know that the time ordered products
$T(\varphi^{k_1} \dots \varphi^{k_n})$ have order $\sum_i k_i$. 
It is also clear that the order is additive under the 
multiplication of two operators. Using the notion of the order of an operator, 
we now give a recursive definition of local, covariant field with ``almost homogeneous scaling''.
\begin{defn}
\label{ahsc}
A local, covariant field $\Phi$ of order zero (i.e., a local c-number field) is 
said to have ``almost homogeneous scaling'' if it scales in fact 
exactly homogeneously, 
\begin{eqnarray}
\lambda^{-d_\Phi}S_\lambda \Phi = \Phi. 
\end{eqnarray}
A local, covariant field $\Phi$ of order $k > 0$ is said to scale almost homogeneously if 
\begin{eqnarray}
\label{dS}
\lambda^{-d_\Phi}S_\lambda\Phi = \Phi + \sum_i
\ln^i \lambda \cdot \Psi_i, \quad \text{for all $\lambda > 0$,}
\end{eqnarray}
where the $\Psi_i$ are finitely many local, covariant fields of order $\le k-1$
with $d_{\Psi_i} = d_\Phi$ and almost homogeneous scaling.
\end{defn}
Our requirement concerning the scaling of local Wick-products and time ordered products is 
then the following. 

\medskip
\noindent
{\bf Scaling:} 
The local time ordered products $\Phi = T(\varphi^{k_1} \dots  \varphi^{k_n})$ 
have almost homogeneous scaling with $d_{\Phi} = \sum k_i =$ order of $\Phi$.

\section{Analysis of the renormalization ambiguity for local Wick products and 
their time ordered products}
\label{sec5}

\subsection{Uniqueness of local Wick products}
We now analyze the ambiguity in defining local Wick powers with the properties stated in the previous 
section. As previously mentioned, we will explicitly consider only undifferentiated 
Wick powers here, but our results can be straightforwardly extended to differentiated 
Wick powers (modulo the remark in section \ref{sec3.0} above). 

\begin{thm}
\label{wuniq}
Suppose we are given two sets of local Wick products $\varphi^k(x)$ and 
$\widetilde \varphi^k(x)$, satisfying the requirements 
formulated in the previous section (for all $k$). Then there holds
\begin{eqnarray}
\label{expan}
\widetilde \varphi^{k}(x) = \varphi^k(x) + 
\sum_{i=0}^{k-2} 
{k \choose i}
C_{k-i}(x) \varphi^i(x).
\end{eqnarray} 
Here, 
\begin{equation}
C_k(x) \equiv C_k[g_{ab}(x), R_{abcd}(x), \dots, \nabla_{(e_1} \dots \nabla_{e_{k-2})}
R_{abcd}(x), \xi, m^2]
\quad (k \in \mn)
\end{equation} 
are polynomials (with real coefficients depending analytically on $\xi$)
in the metric, the curvature and the mass parameter, 
which scale as $C_k \to \lambda^k C_k$ under rescalings $g_{ab} \to \lambda^{-2}g_{ab}$, 
$m^2 \to \lambda^2 m^2, \xi \to \xi$.
\end{thm}
\noindent
{\it Remark:} The space of possible curvature terms $C_k$ described in the theorem is 
finite dimensional for every $k$. For example $C_2$ must be a real linear combination of 
$R$ and $m^2$, since these are the only curvature terms with the required properties.
Therefore the ambiguity in defining $\varphi^2$ is given by 
$
\widetilde \varphi^2  = \varphi^2 + (Z_1 R + Z_2 m^2)1, 
$
where $Z_1, Z_2$ are undetermined real constants, depending analytically on $\xi$. 

\medskip
\noindent
{\it Proof of Thm. \ref{wuniq}:}
The proof is divided into two steps: We first show that there exist local, 
covariant, Hermitian c-number fields $C_k$ such that Eq. \eqref{expan} holds and which have the property that
each $C_k$ depends continuously (analytically) on the metric and 
scales homogeneously up to logarithmic terms with dimension $d_{C_k} = k$.  
The second step is then to show that the $C_k$ are
polynomials in the metric, the Riemann tensor, its derivatives 
and the coupling constants, and that they  
scale in fact exactly as $C_k \to \lambda^{k} C_k$ under
a rescaling of the metric and the mass parameter. 

The first step is accomplished by a simple induction
argument in $k$. Clearly, Eq.~\eqref{expan} holds for $k=1$ and $C_1 = 0$, since 
there is no ambiguity in the definition of the free field. Suppose we have found 
Hermitian local c-number fields $C_i$, $i=2, 3, \dots, k-1$ such that Eq. \eqref{expan}
holds up to order $k-1$ and which have furthermore the properties (a) they are 
continuous (analytic) under corresponding variations of the metric and the 
parameters and (b) they have almost homogeneous scaling with dimension $d_{C_i} = i$. 
We define a local, covariant field $\Phi_k$ by 
\begin{eqnarray}
\label{phikdef}
\Phi_k(x) \mydef \widetilde \varphi^k(x)  -  \left(\varphi^k(x) +  \sum_{i=1}^{k-2} 
{k \choose i}C_{k-i}(x) \varphi^i(x)\right). 
\end{eqnarray}
By the induction assumption it follows that 
the local, covariant field $\Phi_k$ is Hermitian, it is continuous
(analytic) under corresponding variations of the metric and 
the parameters, and it has almost homogeneous scaling with $d_{\Phi_k} = k$. 
This is because $\Phi_k$ arises as a sum of of local, covariant fields 
with these properties. Using the expansion requirement for the 
local Wick powers and the inductive assumption, one easily gets
\begin{equation}
[\Phi_k(x), \varphi(y)] = 0 \quad \text{for all $x, y \in M$.} 
\end{equation}
Using Prop. \ref{pro1} we therefore get that $\Phi_k = C_k 1$, where 
$C_k \equiv C_k[\g, p]$ is some Hermitian local, covariant 
c-number field with the 
properties (a) and (b). Using the microlocal spectrum condition for the 
local Wick monomials, we moreover immediately get that $C_k$ is 
actually a smooth function in $x$. 
We have thus completed the first step and we come to the second step.

The locality requirement, Def. \ref{2.1}, implies that\footnote{Note that the 
role played by $\iota_\chi$ in the locality requirement is trivial in the case at hand, since
$C_k$ is a c-number.}
\begin{eqnarray}
\label{cloc}
\chi^*C_k[\g, p] = C_k[\chi^* \g, p], 
\end{eqnarray}
for any diffeomorphism $\chi$ of $M$, and that $C_k[\g, p](x) = C_k[\g', p](x)$
holds true whenever $\g = \g'$ in some open neighborhood of the point $x$. The first 
condition means that $C_k[\g, p](x)$ is given by a diffeomorphism covariant expression, and 
the second means that it depends only on the germ of $\g$ at $x$.
In order to proceed, we now consider the subspace of all metrics $\g$, which are 
real analytic in some neighborhood of $x$, and we view 
$C_k$ as a functional on that sub-space. Since the germ at $x$ of a real 
analytic metric $\g$ depends only on the metric itself and all its derivatives at $x$, 
this functional must be of the form   
\begin{eqnarray}
C_k[\g, p](x) \equiv C_k[g_{\mu\nu}(x), \stackrel{\circ}{\partial}_\sigma 
g_{\mu\nu}(x), \stackrel{\circ}{\partial}_\sigma \stackrel{\circ}{\partial}_\rho g_{\mu\nu}(x), 
\dots, p]
\end{eqnarray}
for all real analytic metrics $\g$. Here, $\stackrel{\circ}{\partial}_\mu$ is 
the coordinate derivative operator in some fixed analytic coordinate system 
around $x$ and greek indices denote the components in these coordinates. 
For convenience, we take the values of all the coordinates of $x$
to be zero. Consider, now, the 1-parameter family of coupling 
parameters $p^{(s)} = (\xi, s^2 m^2)$ and the following 1-parameter 
family of real analytic metrics, defined by 
\begin{eqnarray}
\label{gsdef}
\g^{(s)} = s^{-2} \chi_s^* \g. 
\end{eqnarray}
Here, $\chi_s$ is the diffeomorphism 
which in our coordinates around $x$ acts by rescaling the 
coordinates by a factor $s$. Let $y^\alpha$ denote the 
coordinates of a point $y$ in a sufficiently small neighborhood 
of $x$. In terms of components in our fixed coordinate system, we 
have 
\begin{equation}
\label{*}
g^{(s)}_{\mu\nu}(y^\alpha) = g_{\mu\nu}^{\,}(sy^\alpha).  
\end{equation}
It follows immediately from \eqref{*} that $\g^{(s)}$ is an analytic family of metrics in a 
neighborhood of $x$ and $s=0$. 
By the analyticity and analytic microlocal scaling degree requirements, 
$C_k[\g^{(s)}, p^{(s)}](x)$ is analytic in 
$s$ in a neighborhood of $s = 0$, and we may thus expand it in a convergent power 
series about $s = 0$. 
It also follows immediately from \eqref{*} that 
$\stackrel{\circ}{\partial}_{\sigma_1} \dots \stackrel{\circ}{\partial}_{\sigma_k}  
g_{\mu\nu}^{(0)}(y) = 0$ for all $y$ in a neighborhood $x$ and that
\begin{eqnarray}
g_{\mu\nu}^{(s)}(x) = g_{\mu\nu}^{\,}(x)\quad 
\myn_{\sigma_1} \dots \myn_{\sigma_k}  
g_{\mu\nu}^{(s)}(x) = s^k \myn_{\sigma_1} \dots \myn_{\sigma_k}  
g_{\mu\nu}^{\,}(x).
\end{eqnarray}
We find from this the power series expansion
\begin{multline}
C_k[\g^{(s)}, p^{(s)}](x) = 
\sum_{n = 0}^\infty s^n \sum_{2j_0 + j_1 + 2j_2 + \dots + rj_r = n}
\frac{\partial^{j_0 + j_1 + \dots + j_r} 
C_k[\dots]}{(\partial m^2)^{j_0} [\partial( \myn\!\g(x))]^{j_1} \dots
[\partial (\myn \dots \myn\!\g(x))]^{j_r}} \\ \times
m^{2j_0}[(\myn \!\g)(x)]^{j_1} \dots [(\myn \dots \myn\!\g)(x)]^{j_r},
\end{multline}
where the spacetime indices have been omitted for simplicity and where 
$$
[\dots] = [g_{\mu\nu}(x), 0, \dots, 0, \xi, m^2 = 0].
$$
Applying Eq. \eqref{cloc} to the diffeomorphism $\chi_s$ and using that $\chi_s(x) = x$, we get 
\begin{equation}
C_k[\g^{(s)}, p^{(s)}](x) = C_k[s^{-2}\g, \xi, s^2 m^2](x). 
\end{equation}
Let us define  
$K_n[g_{\mu\nu}(x), \dots, \myn_{\sigma_1} \dots \myn_{\sigma_n} g_{\mu\nu}(x), \xi, m^2]$ 
(which we shall simply denote by $K_n[\g, \xi, m^2](x)$) 
as the coefficient of $s^n$ in the above power series expansion, 
\begin{equation}
\label{cexp}
C_k[s^{-2}\g, \xi, s^2 m^2](x) \equiv \sum_{n = 0}^\infty s^n K_n[\g, \xi, m^2](x).   
\end{equation}
(Note that $K_n$ is a polynomial in $m^2$ and the derivatives of the metric, 
whose coefficients depend analytically on $\xi$.)
The left side of this identity is covariant under diffeomorphisms for all $s$. Therefore 
it follows that also each individual term in the series on the right side of this 
equation must have this property, i.e., for any analytic diffeomorphism $\chi$,  
\begin{eqnarray}
\label{kcov}
\chi^* K_n[\g, \xi, m^2] = K_n[\chi^* \g, \xi, m^2] \quad \text{for all $n \ge 0$.} 
\end{eqnarray} 
Since $K_n[\g, \xi, m^2](x)$ depends in addition 
polynomially on the metric and its derivatives at $x$, for all $x \in M$, 
it follows from the ``Thomas replacement theorem'' (see \cite[Lem. 2.1]{iw})
that $K_n[\g, \xi, m^2]$ can be written in a ``{\it manifestly} covariant form'', i.e., as 
a polynomial in the metric, the Riemann tensor, a finite number of its 
(symmetrized) metric derivatives and $m^2$, 
whose coefficients depend analytically on $\xi$. In other words
\begin{eqnarray}
K_n[\g, \xi,  m^2](x) \equiv K_n[g_{ab}(x), R_{abcd}(x), \dots, \nabla_{(e_1}
\dots \nabla_{e_{n-2})} R_{abcd}(x), \xi, m^2]. 
\end{eqnarray}

We now use the scaling properties of $C_k$ to find out more 
about its functional dependence on the metric and the coupling
parameters. First, since the scaling dimension of $C_k$ is $k$, 
we immediately find that $K_n = 0$ for all $n < k$.
By Eq. \eqref{cexp}, this means that the map $\lambda \to 
\lambda^{-k}C_k[\lambda^{-2}\g, \xi, \lambda^2 m^2](x)$ is analytic at
$\lambda = 0$. Furthermore, we know that $C_k$ is a local, covariant 
field which scales 
almost homogeneously. This means by definition that 
\begin{equation}
\lambda^{-k} C_k[\lambda^{-2}\g, p(\lambda)] - C_k[\g, p]
= \sum_i \ln^i \lambda \cdot \Psi_i[\g, p], \quad \text{with $p(\lambda) = (\xi, \lambda^2 m^2)$,} 
\end{equation}
for a finite number of local, covariant fields $\Psi_i$. Since the left side 
of this equation is analytic at 
$\lambda = 0$ and since the logarithms are not, this is only possible if 
in fact $\Psi_i = 0$ for all $i$. Therefore, only the $k$-th term in the 
series \eqref{cexp} can be nonzero, which means that 
\begin{eqnarray}
\label{cpoly}
C_k[\g, \xi, m^2](x) \equiv K_k[g_{ab}(x), R_{abcd}(x), \dots, \nabla_{(e_1}
\dots \nabla_{e_{k-2})} R_{abcd}(x), \xi, m^2], 
\end{eqnarray} 
for all analytic metrics $\g$, that is, $C_k$ is a polynomial 
in the metric, the curvature and 
the mass parameter, whose coefficients depend analytically on $\xi$. Since we 
already know that $C_k$ is Hermitian, the coefficients of this polynomial 
must be real. Moreover, we can directly read off from the expansion 
\eqref{cexp} that 
\begin{equation}
C_k[\lambda^{-2}\g, \xi, \lambda^2 m^2](x) = \lambda^{k}C_k[\g, \xi, m^2]. 
\end{equation}
This then proves the theorem for analytic metrics $\g$. But we already know that 
$C_k[\g, p]$ has a continuous dependence on the metric. By 
approximating a smooth metric by a sequence of metrics which are real 
analytic in a neighborhood of $x$, we thus conclude that 
Eq. \eqref{cpoly} must also hold for 
metrics which are only smooth, thus proving the theorem. 
\qed

\subsection{Existence of local Wick products}
\label{sec5.2}
We next sketch how to construct local Wick powers with the desired properties. The construction
is very similar to the construction for the renormalized stress energy 
operator given in \cite{w1}. The main ingredient in our construction is the local
``Hadamard parametrix'', given by Eq. \eqref{Hdef}. $H$ is not defined 
globally but only for $x, y$ contained in a sufficiently small 
convex normal neighborhood.\footnote{The reason for considering convex normal neighborhoods 
is that even $\sigma$ is only defined for points that can be joined by a unique geodesic.}
In the following we therefore restrict attention to such a neighborhood in all 
expressions involving $H$. (This does not create any problems
for our construction of local Wick powers, since only coincident
limits of quantities involving $H$ need to be considered.) 
A technical complication arises from the fact that, 
while $u$ is (at least locally) unambiguously defined for 
arbitrary smooth spacetimes, the same does not apply to $v$, 
which is unambiguously defined only 
for real analytic spacetimes. In the latter case, $v$ is expandable as 
\begin{equation}
\label{**}
v(x, y) = \sum_{n=0}^\infty v_n(x, y) \sigma^n, 
\end{equation}
where $v_n$ are certain real and symmetric \cite{m} smooth functions 
constructed from the metric and $\xi, m^2$. In principle, one would like to 
define $v$ by the above formula also for spacetimes which are only smooth. 
However, it is well-known that the above series does not in general converge in this case. 
This difficulty can be overcome by replacing the coefficients $v_n(x, y)$ in 
the above expansion by $v_n(x, y)\psi(\sigma/\alpha_n)$, where $\psi: \mr \to \mr$ is some smooth
function with $\psi(x) \equiv 1$ for $|x| < \frac{1}{2}$ and $\psi(x) \equiv 0$ for $|x| > 1$.
If the $\alpha_n$'s tend to zero sufficiently fast, then the series with the above modified 
coefficients converges to a smooth function $V$. The coincidence limit of $V$ and of all its
derivatives does not depend on the choice of $\alpha_n$ and $\psi$, and it 
is only through these that $V$ enters our definition of local Wick products.
These choices therefore do not affect our definition.
  
We choose a quasi-free state $\omega$ on $\cA(M, \g)$ and represent $\cW(M, \g)$ as
operators in the GNS representation of $\omega$. Next, we define operator-valued
distributions $\lno \varphi(x_1) \dots \varphi(x_n) \rno_H$ by a 
formula identical to Eq. \eqref{wndef}, except that $\omega$ is replaced by $H$
in that formula. Now, by the very definition of Hadamard 
states,  $H$ is equal, modulo a smooth function,  to the symmetrized two-point function of
$\omega$. Consequently, it follows immediately that
$\lno \varphi(x_1) \dots \varphi(x_n) \rno_H$ 
can be smeared with distributions $t \in \cE'_n(M, \g)$ (supported sufficiently
close to the total diagonal in $M^n$), and the so-obtained expressions
belong to $\cW(M, \g)$. By analogy with our definition of a normal ordered 
field operator, Eq. \eqref{nodef}, we are thus allowed to define
\begin{equation}
\label{phih}
\lno \varphi^k (f) \rno_H \; \mydef \; \int_{M^k} 
\lno \varphi(x_1) \dots \varphi(x_k) \rno_H  f(x_1) 
\delta_\g(x_1, \dots, x_k) \prod_i \mu_\g (x_i).   
\end{equation} 
Although it will not be needed until the next subsection, we find it convenient to define, 
by analogy with Eq. \eqref{nodef1}, also multi-local Wick products of the  
form $\lno \varphi^{k_1}(f_1) \dots \varphi^{k_n}(f_n) \rno_H$. Local Wick 
products involving derivatives of the field can also be defined in a 
similar manner, although, as previously mentioned in the remark in Sec. \ref{sec3.0}, 
the definition fails to satisfy an additional condition that one 
may want to impose. 

\medskip

We claim that the fields $\lno \varphi^k \rno_H$
are local Wick monomials in the sense of the criteria given in Secs. 
\ref{sec2} and \ref{sec3}. We will not 
give a detailed proof of this claim here but merely indicate the main arguments. That 
$\lno \varphi^k \rno_H$ is a local, covariant field immediately follows
from the fact that the Hadamard parametrix is locally and covariantly defined in 
terms of the metric. The expansion
property can be seen in just the same way as the corresponding property for normal ordered
Wick monomials. It seems clear that the construction yields continuous (analytical) dependence of our 
Wick monomials under corresponding variations of the metric and the parameters, although
we have not attempted to give a complete proof of this result. 
Finally, in order to verify the scaling axiom, 
we first restrict our attention to real analytic 
spacetimes $(M, \g)$, so that the function $v$, Eq. \eqref{**}, 
is well-defined.
In that case one finds from the definition of $u$ and $v$ that
\begin{equation}
\label{hscl}
\lambda^{-2} H[\lambda^{-2} \g, \xi, \lambda^2 m^2] = H[\g, \xi, m^2] + v[\g, \xi, m^2] \ln \lambda^2.
\end{equation}
The appearance of the $v\ln\lambda^2$ term is due to the fact that the definition of 
$H$ implicitly depends on a choice of length scale in the argument of the logarithm.\footnote{This
becomes more apparent by writing the logarithmic term in $H$
as $v \ln \sigma \mu^2$, where $\mu$ has 
the dimension of a mass.}
Using Eq. \eqref{hscl} and the definition of the scaling map $S_\lambda$, Eq. \eqref{Slambda}, we find
that $\lno \varphi^{k}\rno_H$ has dimension $k$ and that it scales almost homogeneously in 
the sense of Def. \ref{ahsc}. 
The same holds also for smooth spacetimes, 
by the continuity of the local Wick monomials. Thus we have demonstrated existence of
local Wick products satisfying all of our requirements. 

Although $\lno \varphi^k \rno_H$ scales almost homogeneously, it should be noted 
that the presence of the $\ln \lambda^2$
term in Eq. \eqref{hscl} implies that it fails to scale exactly
homogeneously. The local, covariant fields $\Psi_i$ in Eq. \eqref{dS} 
are given by lower order local Wick monomials times curvature terms of the appropriate
dimension. 
Now, by Eq. \eqref{expan}, any other prescription for the local Wick products, $\varphi^k$, 
will be related to $\lno \varphi^k \rno_H$ by 
\begin{equation}
\varphi^k(x) = \;\lno \varphi^k(x) \rno_H + \sum_{i=0}^{k-2} {k \choose i} C_i(x)
\lno \varphi^i(x) \rno_H
\end{equation}
where each $C_i$ scales exactly homogeneously. It follows that $\varphi^k$ also fails
to scale exactly homogeneously. Consequently, by an argument given on pp.~98--99 of 
\cite{w1}, there is an inherent ambiguity in the definition of $\varphi^k$ that 
cannot be removed within the context of quantum field theory in curved spacetime. Thus, 
in quantum field theory in curved spacetime, the renormalization ambiguities 
arise not only from the definition of the time ordered products of Wick polynomials, 
but also from the local Wick polynomials themselves.

\subsection{Uniqueness of local time ordered products}
\label{mainpar}

The analysis of the ambiguity in the definition of local time ordered products of 
local Wick monomials differs less in substance than in combinatorical 
complexity from the corresponding analysis for the local Wick products. Since 
the combinatorical side is rather well-known, we only sketch the proof of the
result, Thm. \ref{tuniq}. The presentation as well as the proof
of our result is simplified by comparing an arbitrary prescription for 
the time ordered products to a prescription 
based on the local Wick products $\lno \varphi^k(x) \rno_H$, 
defined in the previous subsection. 

Again, for notational simplicity, we explicitly consider only time ordered products
of undifferentiated local Wick products, but our arguments and results would  
apply to time ordered products of differentiated Wick products as well (modulo the 
remark of Sec. \ref{sec3.0}). 

We find it convenient to use a multi-index notation, i.e. $k \in \mn^n$ means a multi-index 
$k = (k_1, \dots, k_n)$, and standard abbreviations for multi-indices 
such as  ${k \choose i}= \prod_j \frac{k_j!}{i_j!(k_j - i_j)!}$. 
$\cP = I_1 \uplus \dots \uplus I_s$ denotes a collection of 
pairwise disjoint subsets of $\{1, \dots, n\}$.  
\begin{thm}
\label{tuniq}
Consider a prescription $T(\prod_i \lno \varphi^{k_i}(x_i)\rno_H)$ 
for defining local time ordered products based on the local Wick products $\lno \varphi^k(x) \rno_H$, 
and another prescription, $\widetilde T(\prod_i \varphi^{k_i}(x_i))$, based 
on another, arbitrary prescription $\varphi^k(x)$ for defining local Wick products. 
Assume that both prescriptions for defining
local time ordered products satisfy all the requirements of Sec. \ref{sec3}. Then  
\begin{multline}
\label{master}
{\widetilde T}\left(\prod_{i=1}^n \varphi^{k_i}(x_i)\right)
\;=\; T\left(\prod_{i=1}^n \lno \varphi^{k_i}(x_i) \rno_H\right) \;\;+ \\  
+ \sum_{
\scriptsize{
\begin{matrix}
\cP = I_1  \uplus \dots  \uplus I_s \\
\text{not all $I_j = \emptyset$}
\end{matrix}
}
}
T\left({\prod_{\scriptsize{I = \{i_1, \dots, i_{|I|}\} \in \cP}}} 
\lno {\cO_{k_I}}
(x_I) \rno_H 
\prod_{i \notin
I\,\forall I \in \cP} \lno \varphi^{k_i}(x_i) \rno_H\right), 
\end{multline}
where $x_I = (x_{i_1}, \dots, x_{i_{|I|}})$
and $k_I = (k_{i_1}, \dots, k_{i_{|I|}})$.
For $n \ge 2$, the $\lno \cO_k(x_1, \dots, x_n) \rno_H$ ($k \in \mn^n$) 
are local, covariant quantum fields of the form 
\begin{eqnarray}
\label{xxx}
\lno {\cO}_k(x_1, \dots, x_n) \rno_H \; \equiv 
\sum_{i \le k} 
\left(
\begin{matrix}
k\\
i
\end{matrix}
\right)
{C}_{k-i}(x_1) 
\delta(x_1, \dots, x_n) \lno \varphi^{i_1}(x_1) \dots \varphi^{i_n}(x_n) \rno_H
\end{eqnarray}
where the ${C}_k$ are real c-number polynomials in $g_{ab}, R_{abcd}, \dots, 
\nabla_{(e_1} 
\dots \nabla_{e_{d-2})} R_{abcd}, m^2$, and covariant derivative 
operators $\nabla^{x_i}_a$, 
with scaling ($=$ engineering) dimension $d = \sum k_i - 4(n-1)$, whose 
coefficients depend analytically on $\xi$. 
For $n = 1$, the quantum fields $\lno {\cO}_k(x) \rno_H$ ($k \in \mn$) 
are given by the same kind of expression as above, but 
with no delta-functions and no covariant derivatives. 
\end{thm}
\noindent
{\it Remarks:} 
(1) 
The multi-local covariant quantum fields $\lno \cO_k(x_1, \dots, x_n) \rno_H$ can
alternatively be written as a sum of, 
possibly differentiated, mono-local Wick powers (i.e., depending only on 
one argument, say, the point $x_1$), multiplied by suitable differentiated delta-functions. 
In formulas, with $(a)$ denoting a $4n$-dimensional spacetime multi index, 
\begin{eqnarray}
\label{omono}
\lno \cO_k(x_1, \dots, x_n) \rno_H \;=\; 
\sum_{(a)} \lno \cC_k^{(a)}(x_1) \rno_H \nabla^{x_1}_{a_1} 
\dots \nabla^{x_n}_{a_n} \delta(x_1, \dots, x_n),   
\end{eqnarray}
where the $\lno \cC_k^{(a)} \rno_H$ are local Wick polynomials, possibly with derivatives
(all spacetime indices are assumed to be raised), 
whose coefficients are polynomials in the metric, the curvature, its
covariant derivatives and the mass. These polynomials scale almost
homogeneously with dimension $\sum_i k_i - 4(n-1)$.

The time ordered products appearing in the second line of Eq. \eqref{master} are 
to be understood as the expressions obtained by inserting the above expression
for the fields $\lno \cO_k \rno_H$ and by pulling the delta function 
type terms out of the time ordered product. 
The disadvantage of writing Eq. \eqref{master} explicitly in terms 
of these monolocal Wick-powers is that the relation between the ambiguities for 
different $k$ and fixed order $n$ (due to the expansion property of the time 
ordered products) now becomes a rather 
complicated-looking constraint on the possible delta-function type terms.
A formulation of Thm. \ref{tuniq} not involving the specific prescription 
$\lno \varphi^k(x)\rno_H$, but instead some other arbitrary prescription, 
would consist in writing all the generalized multilocal Wick products 
in expression \eqref{master} in terms of ordinary, monolocal ones, 
and then replacing these by that arbitrary prescription for those fields.

(2) The collection of local, covariant fields $\lno {\cO}_k(x_1, \dots, x_i) \rno_H$ with $i \le n$ represent the 
finite renormalization ambiguity in defining 
time ordered products with $n$ factors. The crucial point of 
the theorem is that the form of these ambiguities is severely restricted. 
Our uniqueness result for the Wick monomials, Thm. \ref{wuniq}, is a
special case of the above theorem, corresponding to $n=1$.

\medskip
\noindent
{\it Sketch of the proof for Thm. \ref{tuniq}:} 
One proceeds by a double induction in the order $n$ in perturbation 
theory and the scaling dimension $d = \sum k_i$ of the time ordered products. Assuming the validity of 
the theorem up to order $n-1$, one finds, using the causal factorization of the time ordered
products, that Eq. \eqref{master} also holds at order $n$, 
up to an unknown local, covariant $\Phi_k(x_1, \dots, x_n)$ which is nonzero only 
for points such that $x_1 = \dots = x_n$. Assuming
now that this field has the form Eq. \eqref{xxx} for all multi indices $k$ with $\sum k_i \le d-1$, 
one finds that it also has this form for dimension $d$, up to a 
c-number field of 
the form $c_k(x_1, \dots, x_n) = C_k(x_1)\delta(x_1, \dots, x_n)$, where $C_k$ 
is a polynomial in the covariant derivative operators with bounded coefficients. 
By locality, $C_k$ is locally constructed out of the metric and out of the coupling 
parameters. 
The task is then to show that it can be written as a polynomial in 
$g_{ab}, R_{abcd}, \dots, m^2, \nabla^{x_i}_a$, whose coefficients are analytic functions in 
$\xi$, and which scale as $C_k \to \lambda^d C_k$ under 
a corresponding rescaling of the parameters.  

In order to find out more about the functional dependence of $C_k$ on 
the metric, we now use the continuous and analytic dependence of the time ordered products 
under corresponding variations of the metric and the parameters, and their scaling behavior. 
This is done in essentially the same way 
as in our uniqueness proof for the local Wick products, so we only sketch the main
arguments here, focusing on the differences compared to the case of the Wick monomials.
For simplicity, let us first assume that $C_k$ contains no derivatives. 
Consider an analytic family, $\g^{(s)}$, of analytic metrics 
in a neighborhood $O$ in $M$, and an analytic family, $p^{(s)}$, of coupling 
parameters. We would like to show that the distribution 
$C_k^{(s)}(x)$ is analytic in $s$ and $x$. (Here and in the following, the superscript 
$s$ indicates that we mean the quantity associated with the metric $\g^{(s)}$ and 
the coupling parameters $p^{(s)}$.) 
In order to show this, we look at the analytic 
wave front set of $c_k^{(s)}(x_1, \dots, x_n)$, viewed as a distribution 
jointly in $s$ and $x_1, \dots, x_n$. Now, this distribution arises as 
a sum of products of distributions of the form $c_j^{(s)}(x_1, \dots, x_m)$, 
with $m \le n-1$ and $j = (j_1, \dots, j_m)$, and of time ordered products, 
$T^{(s)}(\dots)$. The analytic wave front sets of the $c^{(s)}_j$
(viewed as distributions in $s$ and the $x$-variables) is known by 
the inductive assumption; it is has the same form as the wave front 
set of a delta-distribution. The analytic wave front set of the 
time ordered products---or rather of their expectation value in
some analytic family of states, viewed as a distribution in $s$ and 
the $x$-variables---is known by the analyticity requirement combined 
with the analytic microlocal spectrum condition.  
One can use this information to infer that $c_k^{(s)}(x_1, 
\dots, x_n)$ (viewed as a distribution in $s$ and the $x$-variables)
has analytic wave front set
\begin{multline}
\WF_A(c_k) \subset \{(x_1, p_1, \dots, x_n, p_n, s, \rho) \in T^*(O^n  
\times (-\epsilon, \epsilon)) \backslash \{0\}
\mid\\
(x_1, p_1, \dots, x_n, p_n) \in \Gamma^T_A(O, \g^{(s)}) \},  
\end{multline}
where the conic set $\Gamma^T_A(O, \g^{(s)})$ is specified in 
the analytic microlocal spectrum condition.
But we already know that $c_k$ has support on the set of points 
such that $x_1 = \dots = x_n$. Using this, we therefore find  
\begin{multline}
\WF_A(c_k) \subset \{(x_1, p_1, \dots, x_n, p_n, s, \rho) \in 
T^*(O^n 
\times (-\epsilon,\epsilon)) 
\backslash \{0\} \mid \\ 
x_1 = \dots = x_n, \sum_i p_i = 0, \;\text{not all $p_i = 0$} \}. 
\end{multline} 
Now, we can trivially write 
\begin{equation}
\label{trwr}
C_k^{(s)}(x) = 
\int_{M^{n-1}} c_k^{(s)}(x, y_1, \dots, y_{n-1}) f(y_1, \dots, y_{n-1}) 
\prod_{i=1}^{n-1} \mu^{(s)}(y_i), 
\end{equation}
where $f \in \cD(O)$ is equal to one near $x$. By \cite[Thm. 8.5.4']{h} we can conclude from this that 
that $C_k^{(s)}(x)$---viewed as a distribution jointly in $s$ and $x$---has analytic
wave front set
$$
\WF_A(C_k) = \{(x, p, s, \rho) \mid 
(x, p, y_1, 0, \dots, y_{n-1}, 0, s, \rho) \in \WF_A(c_k) \} = \emptyset
$$
near $x$. Since $x$ was arbitrary, this then shows that 
$C_k^{(s)}(x)$ is jointly analytic in $x$ and $s$. We can now proceed 
as in the uniqueness proof for the local Wick products, by considering the 
particular family of metrics $\g^{(s)}$ (defined in \eqref{gsdef}) and parameters
$p^{(s)} = (\xi, s^2 m^2)$, and following through the same steps as there. 
This then shows us that $C_k$ is indeed 
a polynomial in the metric, the curvature and the mass with engineering dimension $d$, 
whose coefficients depend analytically on $\xi$. 
The case when $C_k(x)$ also contains derivatives, $\nabla^{x_i}_a$, 
can be treated essentially in the same way as above. 
The only difference in the argument is that one has to 
consider more general functions $f$ in Eq. \eqref{trwr}. \qed

\medskip
An important direct consequence of Thm. \ref{tuniq} is the 
renormalizability of $\varphi^4$-theory in curved spacetime, i.e., the perturbative
quantum field theory corresponding to the classical theory given by the Lagrangian $\cL_0 + \cL_1$, 
where $\cL_0$ is the free-field Lagrangian in Eq. \eqref{action}, and where 
$\cL_1 = f \varphi^4$. Observables in this interacting quantum field 
theory can be obtained from the $S$-matrix, given by 
\begin{equation}
\label{smatrix}
S(\cL_1) = 1 + \sum_{n \ge 1} \frac{\i^n}{n!} \int_{M^n}
T(\cL_1(x_1) \dots \cL_1(x_n)) \mu_\g(x_1) \dots \mu_\g(x_n), 
\end{equation}   
viewed here as a formal power series in the coupling constant 
$f$. We note that the above integrals would not in 
general make sense if $f$ were taken to 
be a constant, so we instead take it to be an element in 
$\cD(M)$ which is constant in some region, $O$, of spacetime, where 
we wish to define local observables. Choosing $f$ in this way 
makes the series for $S(\cL_1)$ truncated at 
some $N$ an element in $\cW(M, \g)$. 

Now $S(\cL_1)$ clearly depends on what prescription for the local 
time ordered products one choses in \eqref{smatrix}. So consider 
two different prescriptions, $T( \dots )$ and $\widetilde T( \cdots )$, 
for the time ordered products and denote the corresponding $S$-matrices 
by $S(\cL_1)$ and $\widetilde S(\cL_1)$. Now if it were true that
$\widetilde S(\cL_1) = S(\cL_1 + \delta\cL_1)$ for some local, covariant 
field $\delta\cL_1$ which had the same form as the original 
Lagrangian, then the theories based on different prescriptions 
for the time ordered products would actually be equivalent, the 
effect of $\delta\cL_1$ being merely a redefinition 
of the coupling constants of the theory and of the field strength. 
Theories with this property are called 
``renormalizable''. It is well known that $\varphi^4$-theory in 
Minkowski space belongs to this class of theories. 
We now show that Thm.~\ref{tuniq} implies that this is also the case in 
curved spacetime. 

Without loss of generality, we assume that one of the prescriptions for 
the time ordered products, say the ``non--tilda'' one, is based on local normal 
ordering prescription defined in the previous section. 
Since $\lno \cL_1(x) \rno_H \; = f(x)\!\lno \varphi^4(x) \rno_H$, we 
must investigate the possible form of the fields $\lno \cO_k(x_1, 
\dots, x_n) \rno_H$ in the case that all $k_i = 4$, because these
govern the ambiguities in defining the time ordered products appearing 
in Eq. \eqref{smatrix}. Let us define a field
$\lno \delta \cL_1 \rno_H$ by
\begin{equation}
\int_M \lno \delta\cL_1(x)\rno_H \mu_\g(x) 
\;\mydef \; \sum_{n \ge 1} \int_{M^n} \lno \cO_k(x_1, \dots, x_n) \rno_H
\, \prod_{i=1}^n f(x_i) \mu_\g(x_i),  
\end{equation}
where all $k_i = 4$, viewed as a formal power series in $f$. (When this 
series is truncated at some order $N$, the above equation defines a field in $\cW(M, \g)$.) It 
then follows from the properties of the fields $\lno \cO_k \rno_H$ stated in 
Thm. \ref{tuniq} (applied to the case $k_i = 4$), 
and simple dimensional considerations that $\lno \delta \cL_1 \rno_H$ 
is given by 
\begin{multline}
\lno \delta \cL_1 \rno_H  \; = \sum_{n\ge 1} f^n 
\Big[Z_{0, n}\!\lno g^{ab} \nabla_a \varphi \nabla_b \varphi \rno_H 
+ \,(Z_{1, n}  R + Z_{2, n} m^2)\!\lno \varphi^2 \rno_H +\, Z_{3, n}\!\lno \varphi^4 \rno_H + \\
(Z_{4, n} R^2 + Z_{5, n} R_{ab} R^{ab} + Z_{6, n} 
R_{abcd} R^{abcd} + Z_{7, n} \square R + Z_{8, n} m^2 R + Z_{9, n}  
m^4)1\Big] + \dots, 
\end{multline}
where ``dots'' denotes terms containing derivatives of $f$, 
and where $Z_{i, n}$ are real constants.  
One finds from Eq. \eqref{master}, that 
\begin{equation}
\widetilde S(\cL_1) 
= S(\lno \cL_1 \rno_H + \lno \delta \cL_1 \rno_H)
\end{equation}
in the sense of 
formal power series of operators. Now $\lno \delta \cL_1 \rno_H$ has the 
same form as the original Lagrangian, $\lno \cL_0 \rno_H + \lno \cL_1 \rno_H$, apart 
from the terms proportional to the identity operator in the square brackets, 
and apart from the terms involving
the derivatives of $f$. The terms proportional to the identity contribute
only an overall phase to the $S$-matrix and therefore do not affect the definition of 
the interacting quantum fields derived from the $S$-matrix. The 
terms containing derivatives of $f$ vanish in the formal limit when $f \to {\rm const.}$, but 
for non-constant $f$ they do affect the definition of the observables
in the interacting theory. Nevertheless, it can be shown, using the 
arguments given in Sec. 8 of \cite{bf}, that the interacting 
theory obtained from the interaction Lagrangian 
$\lno \cL_1\rno_H + \lno \delta \cL_1 \rno_H$ locally 
(i.e., in the region $O$ where $f$ is constant) does not depend 
on the terms in $\lno \delta \cL_1 \rno_H$
involving derivatives of $f$. 

This then proves renormalizability 
of $\varphi^4$-theory in curved spacetime, provided of course that time ordered products
satisfying our assumptions do indeed exist.

\section{Conclusions and outlook}
We have constructed, for every globally hyperbolic spacetime $(M, \g)$, an algebra 
$\cW(M, \g)$ containing normal ordered Wick products and time ordered products thereof. 
We then gave a notion of what it means for a field in that algebra to be 
``locally constructed out of the metric'' in a covariant manner. Furthermore, we gave notions 
of analytic resp. continuous dependence of a local, covariant field under corresponding
variations of the metric, and we gave a notion of ``essentially homogeneous'' scaling
of a local, covariant field under suitable rescalings of the metric and the parameters of 
the theory. We then axiomatically characterized local Wick polynomials
and local time ordered products by demanding that they satisfy the above 
requirements together with certain other, natural properties expected from 
a reasonable definition of these quantities. The imposition of these requirements
was shown to reduce the ambiguities in defining these quantities to a finite
number of real parameters. The nature of these ambiguities was shown to 
imply the renormalizability of a self-interacting quantum field theory in curved space. 
By an explicit construction, the existence of 
local Wick products with the desired properties was demonstrated.
However, the issue of the existence of local time ordered products is
beyond the scope of this paper and will be treated elsewhere. 

We mention that our 
notion of the scaling of a local, covariant field makes possible
a renormalization group analysis of the quantum observables in the interacting theory
(posed as an open problem in \cite{bf}), i.e. an analysis of the behavior of 
an observable in the interacting theory under a change of scale. 
Namely, the ``action of a renormalization group transformation'' on an observable in the 
interacting theory is implemented in our framework by the scaling map, $S_\lambda$, defined in Eq. \eqref{Slambda}.  
The task is then to analyse the action of this map on observables in the 
interacting theory. Now, the observables in the interacting theory are 
defined in terms of perturbative expressions involving local time ordered products, and hence 
one only has to analyse the action of $S_\lambda$ on the local time ordered products. 
Consider an expression of the form $T_\lambda(\dots) = \lambda^{-d} S_\lambda T(\dots)$, where
$T(\dots)$ is a local time ordered product with scaling dimension $d$. The rescaled
time ordered product $T_\lambda( \dots )$ is in general not equal to the unscaled 
time ordered product. However, by our uniqueness theorem \ref{tuniq}, 
the scaled time ordered products differ from the unscaled ones by well-specified 
renormalization ambiguities, given by certain real parameters (depending on $\lambda$). 
As explained in the previous section, these parameters correspond
to a finite renormalization of the coupling parameters in the theory. 
The action of $S_\lambda$ (i.e., a renormalization group transformation) 
therefore translates directly into a flow of the coupling parameters (and a 
multiplicative rescaling of the field strength). A detailed calculation of 
these can of course only be done based on a concrete prescription for the 
local time ordered products. 

\vspace{1cm}
\noindent
{\bf Acknowledgements:} We wish to thank Klaus Fredenhagen and Bernard Kay
for helpful discussions. This research was supported in part by 
NSF grants PHY95-14726 and PHY00-90138 to the University of Chicago.

\section{Appendix}

It is well known that the regularity properties of 
a distribution $u \in \cD'(\mr^n)$ are in correspondence with the decay properties
of its Fourier transform. This can be made more precise by introducing
the concept of the ``wave front set'' of a distribution \cite{h}, 
which we shall define now. Let $u$ be a distribution of compact support.
We define $\Sigma(u)$ to be the set of all $k \in \mr^n \backslash \{0\}$
which have no conical\footnote{A cone in $\mr^n$ is 
a subset $V$ with the property that if $k \in V$, then 
also $\lambda k \in V$ for all $\lambda > 0$.} 
neighborhood $V$ such that
\begin{eqnarray*}
|\widehat{u}(p)| \le C_N(1 + |p|)^{-N}
\end{eqnarray*}
for all $p \in V$ and all $N = 1, 2, \dots$.
$\Sigma(u)$ may be thought of as describing the ``singular 
directions'' of $u$. The wave front set provides a more detailed
description of the singularities of a distribution by localizing these 
singular directions. If $u \in \cD'(X)$, 
with $X$ an open subset of $\mr^n$, then we define 
$\Sigma_x(u) = \cap_{f} \Sigma(fu)$, where the intersection 
is taken over all $f \in \cD(X)$ such that $f(x) \neq 0$. 
The wave front set of $u$ is now defined as 
\begin{eqnarray*}
\WF(u) \mydef \{ (x, k) \in X \times (\mr^n \backslash \{0\})
\mid k \in \Sigma_x(u) \}. 
\end{eqnarray*}
If $(x, k) \in \WF(u)$, then 
$x$ is a singular point of $u$, i.e., there is no neighborhood of 
$x$ in which $u$ can be written as a smooth function. Conversely, 
if $x$ is a point such that no $(x,k) \in \WF(u)$, then $x$ 
is a regular point. Differentiation 
does not increase the wave front set, $\WF(\partial u) \subset \WF(u)$. 
The wave front 
set of a distribution is an entirely local concept, and it 
can be shown to transform covariantly under a change of coordinates, 
in the sense that $\WF(\chi^* u) = (d\chi)^t \circ \WF(u)$ for any 
diffeomorphism $\chi$.  This makes it possible to define in an 
invariant way the wave front set of distributions $u$ on a 
manifold $X$. The above transformation property then shows that 
$\WF(u)$ is intrinsically a (conic) subset of $T^*X \backslash \{0\}$, where
$T^*X$ denotes the cotangent bundle of $X$, and where $\{0\}$ means
the zero section in $T^*X$.  
(In this paper, $X$ is typically a product manifold $M \times \dots
\times M$.) 

In this paper we often use the notion of the wave front set to ensure that the 
pointwise product of certain distributions exists, or, more generally, 
to ensure that certain linear maps with distributional 
kernel have a well-defined action on certain distributions (cf. Thms. 8.2.10 and 8.2.13
of ref. \cite{h}). 
The above operations with distributions are
not continuous (even if they are well defined) 
in the usual distribution topology. However,  
they are continuous in the so-called ``H\"ormander pseudo topology'', 
which is defined as follows: Let $\Gamma$ be a closed conic 
set\footnote{By this we mean a set of the form $\Gamma = \{
(x, k) \in U \times \mr^n \mid k \in \Gamma_x\}$, 
where $U$ is a closed set and where 
$\Gamma_x$ is a closed cone in $\mr^n$ for all $x \in U$.} 
in $\mr^n \times \mr^n$, and let 
$\cD_\Gamma'(\mr^n)$ be the set of all distributions $u$ on $\mr^n$ with 
$\WF(u) \subset \Gamma$. We say that a sequence $\{u_\alpha\} \subset
\cD'_\Gamma(\mr^n)$ converges to $u$ in the H\"ormander 
pseudo topology if $u_\alpha \to u$ in the usual sense of 
distributions and if, for 
any open neighborhood $O \subset \mr^n$ and any cone $V \subset \mr^n$ 
such that $\Gamma_x \subset V \; \forall x \in O$ and any $f \in \cD(O)$ there holds 
$$
\sup_{k \notin V} |( \widehat{fu_\alpha} - \widehat{fu})(k)|(1 + |k|)^N \to 0
\quad \forall N \in \mn.     
$$
This notion can be generalized in an invariant manner to smooth manifolds $X$, 
where $\Gamma$ is now a closed conic subset of $T^*X$.

\end{document}